\documentclass[aps,prc,amsfonts,reprint,showpacs,showkeys,nofootinbib,twocolumn,preprintnumbers,superscriptaddress]{revtex4-1}

\usepackage{graphicx}

\usepackage{multirow}

\usepackage{times}
\usepackage{everypage}
\usepackage{gensymb}

\newcommand{\ua}{\uparrow}
\newcommand{\da}{\downarrow}

\newcommand{\Ua}{\Uparrow}
\newcommand{\Da}{\Downarrow}

\begin{document}

\title{Measurement of Target and Double-spin Asymmetries for the 
$\vec e\vec p\to e\pi^+ (n)$ Reaction in the Nucleon Resonance Region at Low $Q^2$}


\newcommand*{\ANL}{Argonne National Laboratory, Argonne, Illinois 60439, USA}
\newcommand*{\ASU}{Arizona State University, Tempe, Arizona 85287, USA}
\newcommand*{\CSUDH}{California State University, Dominguez Hills, Carson, California 90747, USA}
\newcommand*{\CANISIUS}{Canisius College, Buffalo, New York 14208, USA}
\newcommand*{\CMU}{Carnegie Mellon University, Pittsburgh, Pennsylvania 15213, USA}
\newcommand*{\CNU}{Christopher Newport University, Newport News, Virginia 23606, USA}
\newcommand*{\CUA}{Catholic University of America, Washington, D.C. 20064, USA}
\newcommand*{\SACLAY}{CEA, Centre de Saclay, Irfu/Service de Physique Nucl\'eaire, 91191 Gif-sur-Yvette, France}
\newcommand*{\UCONN}{University of Connecticut, Storrs, Connecticut 06269, USA}
\newcommand*{\FU}{Fairfield University, Fairfield, Connecticut 06824, USA}
\newcommand*{\FIU}{Florida International University, Miami, Florida 33199, USA}
\newcommand*{\FSU}{Florida State University, Tallahassee, Florida 32306, USA}
\newcommand*{\GWUI}{The George Washington University, Washington, DC 20052, USA}
\newcommand*{\ISU}{Idaho State University, Pocatello, Idaho 83209, USA}
\newcommand*{\INFNFE}{INFN, Sezione di Ferrara, 44100 Ferrara, Italy, USA}
\newcommand*{\INFNFR}{INFN, Laboratori Nazionali di Frascati, 00044 Frascati, Italy}
\newcommand*{\INFNGE}{INFN, Sezione di Genova, 16146 Genova, Italy}
\newcommand*{\INFNRO}{INFN, Sezione di Roma Tor Vergata, 00133 Rome, Italy}
\newcommand*{\INFNTUR}{INFN, Sezione di Torino, 10125 Torino, Italy}
\newcommand*{\ORSAY}{Institut de Physique Nucl\'eaire, CNRS/IN2P3 and Universit\'e Paris Sud, Orsay, France}
\newcommand*{\ITEP}{Institute of Theoretical and Experimental Physics, Moscow, 117259, Russia}
\newcommand*{\JMU}{James Madison University, Harrisonburg, Virginia 22807, USA}
\newcommand*{\KNU}{Kyungpook National University, Daegu 702-701, Republic of Korea}
\newcommand*{\MISS}{Mississippi State University, Mississippi State, Mississippi 39762, USA}
\newcommand*{\UNH}{University of New Hampshire, Durham, New Hampshire 03824, USA}
\newcommand*{\NSU}{Norfolk State University, Norfolk, Virginia 23504, USA}
\newcommand*{\OHIOU}{Ohio University, Athens, Ohio  5701, USA}
\newcommand*{\ODU}{Old Dominion University, Norfolk, Virginia 23529, USA}
\newcommand*{\RPI}{Rensselaer Polytechnic Institute, Troy, New York 12180, USA}
\newcommand*{\URICH}{University of Richmond, Richmond, Virginia 23173, USA}
\newcommand*{\ROMAII}{Universita' di Roma Tor Vergata, 00133 Rome Italy}
\newcommand*{\MSU}{Skobeltsyn Institute of Nuclear Physics, Lomonosov Moscow State University, 119234 Moscow, Russia}
\newcommand*{\SCAROLINA}{University of South Carolina, Columbia, South Carolina 29208, USA}
\newcommand*{\TEMPLE}{Temple University,  Philadelphia, Pennsylvania 19122, USA}
\newcommand*{\JLAB}{Thomas Jefferson National Accelerator Facility, Newport News, Virginia 23606, USA}
\newcommand*{\UTFSM}{Universidad T\'{e}cnica Federico Santa Mar\'{i}a, Casilla 110-V Valpara\'{i}so, Chile}
\newcommand*{\EDINBURGH}{Edinburgh University, Edinburgh EH9 3JZ, United Kingdom}
\newcommand*{\GLASGOW}{University of Glasgow, Glasgow G12 8QQ, United Kingdom}
\newcommand*{\VT}{Virginia Tech, Blacksburg, Virginia 24061, USA}
\newcommand*{\VIRGINIA}{University of Virginia, Charlottesville, Virginia 22901, USA}
\newcommand*{\WM}{College of William and Mary, Williamsburg, Virginia 23187, USA}
\newcommand*{\YEREVAN}{Yerevan Physics Institute, 375036 Yerevan, Armenia}
\newcommand*{\NOWJLAB}{Thomas Jefferson National Accelerator Facility, Newport News, Virginia 23606, USA}
\newcommand*{\NOWLANL}{Los Alamos National Laboratory, Los Alamos, New Mexico 87544, USA}
\newcommand*{\NOWZZZ}{unused, unused}

%
\author{X.~Zheng}
\affiliation{\VIRGINIA}

\author {K.P. ~Adhikari} 
\affiliation{\MISS}
\affiliation{\ODU}

\author{P. Bosted}
\affiliation{\JLAB}

\author{A.~Deur}
\affiliation{\JLAB}
\affiliation{\VIRGINIA}

\author{V. Drozdov}
\affiliation{\INFNGE}
\affiliation{\MSU}

\author {L.~El~Fassi} 
\affiliation{\MISS}
\affiliation{\ANL}


\author{Hyekoo~Kang}
\affiliation{Seoul National University, Seoul, Korea}

\author{K. Kovacs}
\affiliation{\VIRGINIA}

\author{S.~Kuhn}
\affiliation{\ODU}

\author{E.~Long}
\affiliation{\UNH}

% ...

\collaboration{The CLAS Collaboration}
\noaffiliation


\date{\today}%

\begin{abstract}               
We report measurements of target- and double-spin asymmetries for
the exclusive channel $\vec e\vec p\to e\pi^+ (n)$ in the nucleon resonance
region at Jefferson Lab using the CEBAF Large Acceptance Spectrometer (CLAS). 
These asymmetries were extracted from data obtained using a longitudinally polarized NH$_3$ 
target and a longitudinally polarized electron beam with energies 1.1, 1.3, 
2.0, 2.3 and 3.0 GeV. The new results are consistent with previous CLAS
publications but are extended to a low $Q^2$ range from 
$0.0065$ to $0.35$ (GeV$/c$)$^2$.  
The $Q^2$ access was made possible by a custom-built Cherenkov 
detector that allowed the detection of electrons for 
scattering angles as low as $6^\circ$. These results are compared with 
the unitary isobar models JANR and MAID, the partial-wave analysis 
prediction from SAID 
and the dynamic model DMT. In many kinematic regions our results, 
in particular results on the target asymmetry, help to 
constrain the polarization-dependent components of these models. 
\end{abstract}

\pacs{
13.60.Le, 
13.88.+e, 
14.20.Gk 
} 

\maketitle
\section{Physics Motivation}\label{sec:motivation}

%
%
The perturbative nature of the strong interaction at small distances -- often 
referred to as ``asymptotic freedom'' -- was established more than 30 years ago 
and provided strong support for quantum chromodynamics (QCD) to be accepted 
as the correct theory for strong interactions~\cite{Gross:1973id,Politzer:1973fx}. 
On the other hand, calculations at long-distances are still beyond reach because of the
non-perturbative nature at this scale. As a result, we are still far away 
from being able to describe the strong force as it manifests itself in the 
structure of baryons and mesons~\cite{Burkert:2004sk,Aznauryan:2011qj}. 

A fundamental approach to resolve this difficulty is to develop accurate numerical 
simulations of QCD on the lattice; for recent reviews 
see~\cite{Hoelbling:2014uea,Ukawa:2015eka}. 
However lattice QCD methods are difficult to 
apply to light-quark systems such as the nucleon. 
Alternatively, hadron 
models with effective degrees of freedom have been constructed to interpret data. 
One example is the chiral perturbation theory~\cite{Bernard:1993bq,Bernard:1996bi}, 
which is constrained only by the symmetry properties of QCD. 
The constituent quark model, though not fully understood, is one 
successful example that works almost everywhere from hadron spectroscopy to 
deep inelastic scattering~\cite{Zheng:2003un,Zheng:2004ce}. Predictions for the scattering 
amplitudes and polarization-dependent 
asymmetries exist for many resonances within the framework of
the relativistic constituent
quark model (RCQM)~\cite{Warns:1989xr} and the single quark transition model 
(SQTM)~\cite{Burkert:2002zz}.  

The comparison between these predictions
and experimental results, on the other hand, is not straightforward. 
This is because the experimentally measured cross sections and asymmetries
are usually complicated combinations of resonant and non-resonant amplitudes 
and couplings, and their interference terms. To compare with theories, 
partial wave analyses
are often used to extract these amplitudes and resonance couplings from data. 
Once comparisons can be made, data are used to provide inputs for constructing 
or adjusting meson production mechanisms in theories and models, such as 
proper treatment of the hadronic final state and implementation of the 
non-resonant part of the meson production amplitude. These mechanisms are 
usually not included in quark models.
Examples of phenomenological partial wave analyses that can benefit from more data 
are MAID~\cite{Drechsel:2007if}, JANR~\cite{Aznauryan:2009mx}, SAID~\cite{Arndt:2009nv}, 
and the DMT~\cite{Kamalov:1999hs} models. 
Electron-scattering data used to test these calculations include primarily 
$N-N^*$ transition form factors and response 
functions for meson production reactions obtained from Jefferson Lab (JLab), 
MAMI and MIT-Bates. 
Recently, polarization observables such as double spin asymmetries 
and target spin asymmetries
for pion electro-production from the proton have made the beam- and target-helicity response 
functions accessible~\cite{DeVita:2001ue,DeVita:2002uv,Biselli:2003ym,EG1b-Pierce-thesis}, 
providing a new approach to testing models and to a greater understanding of 
the baryon resonance structure.
As an example, the MAID model was based mostly on unpolarized data and is only recently 
being tested extensively against double polarization asymmetries. 
In general, polarization observables provide an important constraint on 
the understanding of the underlying helicity response functions 
or interference terms in $N\to \Delta$ and $N\to N^*$ resonances.

Compared to the proton, existing data on neutron excitation were particularly 
sparse. Neutron data have recently become available from 
JLab~\cite{EG1b-Careccia-thesis,Bosted:2016leu}, which make it 
possible to test the isospin structure of models such as RCQM and SQTM. 
The neutron data will be valuable to the development of 
many phenomenological analyses as well because they need to incorporate 
double polarization asymmetry data for all pion production channels 
from both the proton and the neutron 
to perform the full isospin decomposition. 

In addition, data at very low $Q^2$ values are often desired for testing the chiral
perturbation theory and to study the transition from virtual photons to the real photon point ($Q^2=0$). 
Here, $Q^2$ is defined as $Q^2\equiv -q^2$, 
where $q\equiv(\nu,\vec q)$ is the four-momentum transferred from the incident
electron to the target and 
\begin{eqnarray}
 \nu\equiv E-E'~,\label{eq:omega}
\end{eqnarray}
with $E$ and $E'$ the incident and the scattered electron's energies, respectively. 
At low energy transfers $\nu<2$~GeV the most prominent resonances are the $\Delta(1232)3/2^+$, 
$N(1520)3/2^-$, and $N(1680)5/2^+$~\cite{Warns:1989xr}. For the $N(1520)3/2^-$ and 
$N(1680)5/2^+$, their amplitudes at large $Q^2$ are determined by perturbative QCD
and hadron helicity conservation. It is expected in this region that $A^N\to 1$, where
$A^N$ is the virtual photon helicity asymmetry defined as:
\begin{eqnarray}
  A^N &=& \frac{\vert A_{1/2}\vert^2 - \vert A_{3/2}\vert^2}{\vert A_{1/2}\vert^2 + \vert A_{3/2}\vert^2}~,
\label{eq:AN}
\end{eqnarray}
with $A_{1/2,3/2}$ the scattering amplitudes and the subscripts indicate the
total spin projection of the virtual photon and the nucleon target along the
virtual photon's momentum.
However, data using real photons show
a strong helicity-3/2 dominance and $A^N\to -1$~\cite{PDG}. This indicates that $A^N$
for these two resonances must cross zero at some intermediate $Q^2$ and there have
been calculations for the $Q^2$-dependence of $A^N$ from various 
models~\cite{Warns:1989xr,Burkert:2002zz,Capstick:1992uc}. 
For pion electroproduction, the double spin asymmetry is dominated by
$A^N$~\cite{DeVita:2001ue} and thus data on this observable  
will allow us to test a possible sign flip for the $N(1520)3/2^-$ and $N(1680)5/2^+$ resonances.
Data on the double spin asymmetry of pion photoproduction have recently become available
from the CBELSA/TAPS Collaboration~\cite{Gottschall:2013uha} and are also expeced 
from JLab experiments~\cite{Iwamoto:2012zza,Schott:2015fbn,E04-102}, all used the
frozen spin target with a longitudinal polarization and a circularly polarized photon beam. These
photoproduction data will further test the transition to the real photon point.

\subsection{Formalism for pion electroproduction}\label{sec:intro_formula}

\begin{figure}[!ht]
\begin{center}
\includegraphics[width=0.5\textwidth]{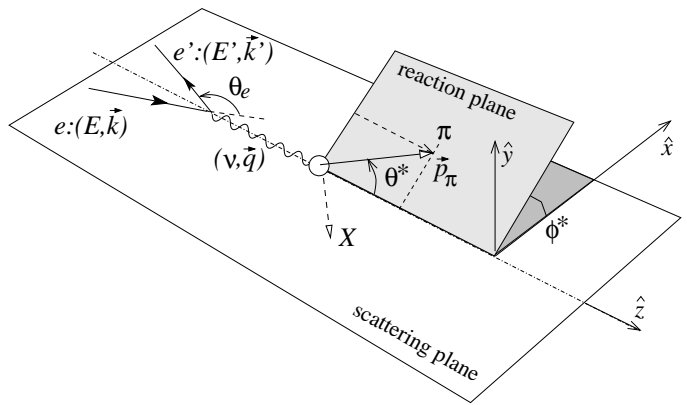}
\end{center}
\caption{Kinematics of single pion electro-production.  The Lorentz boost associated
with the transformation from the laboratory to the CM frame of the $\gamma^*N$ system is
along the momentum transfer $\vec q$, where the coordinates $\hat x, \hat y, \hat z$ 
of the CM frame are defined in this picture.}\label{fig:exclkin}
\end{figure}

Figure~\ref{fig:exclkin} shows the kinematics of single pion
production in the Born approximation: The electron transfers a virtual 
photon $\gamma^*$ of four-momentum $q\equiv (\nu,\vec q)$ to the target nucleon $N$ which forms a 
nucleon resonance. The resonance then decays into a pion and another 
particle $X$. Two planes are used to describe this process:
the scattering (leptonic) plane defined by the incoming and outgoing electrons' 
momenta $\vec k$ and $\vec k^\prime$, and the reaction (hadronic) plane defined 
by the momentum of the virtual photon $\vec q$ and the 
momentum of the outgoing pion $\vec p_\pi$.

The reaction is usually described in terms of $Q^2$, 
the invariant mass $W$ of the $\gamma^*N$ system (which 
is also the $\pi X$ system), and two angles $\theta^*$ and $\phi^*$. 
Here, $\theta^*$ is the angle formed by $\vec q$ and $\vec p_\pi$, 
and $\phi^*$ is the angle formed by rotating the leptonic plane to the hadronic plane. 
If one defines the $\gamma^* N$ center of mass (CM) frame with $\hat z$ pointing along 
$\vec q$, $\hat y$ along $\vec q\times\vec k$, then $\theta^*$ and $\phi^*$ are the 
polar and the azimuthal angles of the emitted pion. 
The energy transfer is related to $Q^2$ and $W$ via 
\begin{eqnarray}
 \nu = \frac{W^2+Q^2-M^2}{2M}~,  \label{eq:omega2}
\end{eqnarray} 
with $M$ the nucleon mass. 
The differential cross section for the reaction $\vec e\vec N\to e\pi(X)$ 
with longitudinally polarized beam and target 
can be written in the following form 
\begin{eqnarray}
\frac{d^5\sigma_h}{dE_{e^\prime} d\Omega_{e^\prime} d\Omega_\pi^*}
&=&\Gamma\frac{d\sigma_h}{d\Omega^*_\pi}~, 
\end{eqnarray}
with
\begin{eqnarray}
\frac{d\sigma_h}{d\Omega^*_\pi}&=& \frac{d\sigma_{0}}{d\Omega_\pi^*}
+P_b\frac{d\sigma_{e}}{d\Omega_\pi^*}
+P_t\frac{d\sigma_{t}}{d\Omega_\pi^*}
+P_b P_t\frac{d\sigma_{et}}{d\Omega_\pi^*} \label{eq:sigma_h}
\end{eqnarray}
where $P_b$ and $P_t$ are respectively the polarizations of the electron beam and the
target along the beam direction, $\sigma_0$ is the unpolarized cross section, and  
$\sigma_e$, $\sigma_{t}$ and $\sigma_{et}$ are the polarized cross section terms 
when beam, target, and both beam and target are polarized. 
Note that the differential cross sections on the right-hand side of 
Eq.~(\ref{eq:sigma_h}) are defined in the CM
frame of the $\gamma^* N$ system, as indicated by the asterisk in the pion's solid angle. 
The virtual photon flux is
\begin{eqnarray}
\Gamma&=&\frac{\alpha k_{\gamma}^\mathrm{lab}}{2\pi^2Q^2}\frac{E^\prime}{E}
\frac{1}{1-\epsilon}~,
\end{eqnarray}
where $\alpha$ is the electromagnetic coupling constant, 
$k_{\gamma}^\mathrm{lab}=(W^2-M^2)/2M$ is the photon 
equivalent energy in the laboratory frame, i.e. the energy needed by a real photon to 
excite the nucleon to an invariant mass $W$. 
The virtual photon polarization is given by 
\begin{eqnarray}
\epsilon&=&
\left[1+\frac{2\vert {\vec q} \vert^2}{Q^2} 
   \tan^2\frac{\theta_e}{2}\right]^{-1} \,, 
\end{eqnarray}
where $\theta_e$ is the angle between the incident and outgoing electrons 
in the laboratory frame.  The $Q^2$ can be calculated as 
\begin{eqnarray}
Q^2 &=& 4 E E^\prime \sin^2 \frac{\theta_e}{2}~.\label{eq:q2}
\end{eqnarray}

To evaluate the pion's kinematics in the CM frame of the $\gamma^* N$ system, 
we relate a laboratory-frame 4-momentum vector $p^\mu$ to the CM-frame
$p^\mu_{cm}$ via a Lorentz boost
with ${\vec \beta } = \hat{z} \vert {\vec q} \vert /(\nu + M)$ and $\gamma=(\nu+M)/W$: 
\begin{eqnarray}
p^0_\mathrm{cm} &=& \gamma p^0 - \gamma\beta p^z ~, \\
p^x_\mathrm{cm} &=& p^x~, \\
p^y_\mathrm{cm} &=& p^y~, \\
p^z_\mathrm{cm} &=&-\gamma\beta p^0 +  \gamma p^z ~.
\end{eqnarray}
Specifically, we have for the virtual photon: 
\begin{eqnarray}
\vert{\vec q}_\mathrm{cm}\vert &=&\frac{M}{W}\vert {\vec q} \vert ~, \\
\nu_\mathrm{cm} &=&\frac{\nu M - Q^2}{W} ~. 
\end{eqnarray}
For the pion
\begin{eqnarray}
  E_{\mathrm{cm},\pi} &=& \gamma\left(E_\pi-\beta\vert\vec p_\pi\vert\cos\theta_\pi\right)~,\\
  p_{z,\mathrm{cm},\pi} &=& \gamma\left(\vert\vec p_\pi\vert\cos\theta_\pi-\beta E_\pi\right)~,
\end{eqnarray}
where $\theta_\pi=\arccos[(\vec q\cdot\vec p_\pi)/(\vert\vec q\vert\vert\vec p_\pi\vert)]$ 
is the angle between the pion momentum and $\vec q$ 
in the laboratory frame, and 
$E_\pi$ is the pion energy again in the laboratory frame. 
The polar angle of the pion in the CM frame is given by 
\begin{eqnarray}
 \theta^*=\arccos\left[\frac{p_{z,\mathrm{cm},\pi}}{\sqrt{E_{\mathrm{cm},\pi}-m_\pi^2}}\right]
\end{eqnarray}
where $m_\pi$ is the pion mass. 
The azimuthal angle of the pion is the same in the laboratory and the CM frame, given by
\begin{eqnarray}
  \phi^* &=& \arccos\left[\frac{\vec a\cdot\vec b}{\vert\vec a\vert\vert\vec b\vert}\right]
\label{eq:phicm}
\end{eqnarray}
with $\vec a\equiv \vec q\times \vec k$ and $\vec b\equiv \vec q\times\vec p_\pi$. In this
paper, the range of $\phi^*$ is defined from $0$ to $2\pi$, i.e. a shift of $2\pi$ is 
added to $\phi^*$ if the result from Eq.~(\ref{eq:phicm}) is negative.

The beam, target and double beam-target asymmetries are
\begin{eqnarray}
 A_{LU} &=& \frac{\sigma_e}{\sigma_{0}}~,\label{eq:ALUdef}\\
 A_{UL} &=& \frac{\sigma_t}{\sigma_{0}}~,\label{eq:AULdef}\\
 A_{LL} &=& - \frac{\sigma_{et}}{\sigma_{0}}~\label{eq:ALLdef}~,
\end{eqnarray}
where each cross section $\sigma$ stands for the $d\sigma/d\Omega_\pi^*$ 
of Eq.~(\ref{eq:sigma_h}). 
Note that we have adopted an extra minus sign in the definition of $A_{LL}$ 
to be consistent with Eq.~(\ref{eq:AN}) and previous CLAS 
publications~\cite{DeVita:2001ue,DeVita:2002uv,Biselli:2003ym}.


In this paper, we report on results of both $A_{UL}$ and $A_{LL}$ extracted 
from the JLab CLAS EG4~\cite{PR03006,PR05111} data. The beam asymmetry
$A_{LU}$ was also extracted from the data, but was used only as a cross-check of the
beam helicity and is not presented here. These results
are available for download from the CLAS database.

\subsection{Previous data}
The first double-spin asymmetry for the $\pi^+n$ channel was published 
based on the CLAS EG1a data with a 2.6 GeV beam, for a $Q^2$ range from 
0.35 to 1.5~(GeV/$c$)$^2$~\cite{DeVita:2001ue,DeVita:2002uv}. 
The $\vec e\vec p\to e'p(\pi^0)$ channel was analyzed for the $\Delta(1232)3/2^+$ 
region using the same dataset~\cite{Biselli:2003ym}. 
Similar analysis using the CLAS EG1b data was 
completed~\cite{EG1b-Pierce-thesis,Bosted:2016leu}, 
in which the target and the double spin asymmetries were extracted 
from both the $\vec e\vec p\to e'\pi^+(n)$ and $\vec e\vec n\to e'\pi^-p$ 
channels using 1.6 -- 5.7~GeV beams with $Q^2$ as low as 0.1~(GeV/$c$)$^2$.

\section{The JLab CLAS EG4 Experiment}
The main physics goal of the CLAS EG4 experiment~\cite{PR03006,PR05111} was 
to measure the inclusive spin structure functions on the proton and the deuteron, and to 
extract the generalized Gerasimov-Drell-Hearn (GDH) sum near the photon point.
The original GDH sum rule~\cite{Gerasimov:1965et,Drell:1966jv}, defined for real 
photons, is a fundamental
prediction on the nucleon's spin structure that relates the helicity-dependent 
total photo-absorption cross section to the nucleon anomalous
magnetic moment. The definition of the GDH sum was generalized to virtual
photons~\cite{Anselmino:1988hn,Ji:2001yu}, and 
the value of the generalized GDH sum at low $Q^2$ was predicted in the
chiral perturbation theory. Similar to the pion production results presented here, 
the goal of the EG4's inclusive analysis is to test the chiral 
perturbation theory prediction and to compare the extrapolation to the $Q^2=0$ point
with the GDH sum rule of the real photon. 
 
The experiment was carried out in 2006 in experimental Hall B of 
JLab.  Inclusive data were collected in the 
range $1<W<2$~GeV/$c^2$ and
$Q^2$ down to $0.015$~(GeV/$c$)$^2$~\cite{EG4-Kang-thesis}, using six beam energies 
(1.1, 1.3, 1.5, 2.0, 2.3, 3.0 GeV) 
on a polarized NH$_3$ target and two energies (1.3, 2.0 GeV) on a polarized ND$_3$
target. 
The average polarizations of NH$_3$ and ND$_3$ 
typically ranged within $(75-90)\%$ and $(30-45)\%$, respectively.
For the exclusive channel, only NH$_3$ data with beam energies of 1.1, 1.3, 
2.0, 2.3, and 3.0 GeV were analyzed with 
the lowest $Q^2$ being $0.0065$~(GeV/$c$)$^2$. The 1.5 GeV energy data were excluded because 
they were taken for
run commissioning purpose and had limited statistics. For ND$_3$ data, the target spin
direction was not flipped during the run, which makes it impossible to extract $A_{UL}$ nor the 
complete information on $A_{LL}$ from the exclusive channel. 

\subsection{The CLAS detector}
\begin{figure}[!ht]
 \includegraphics[width=0.5\textwidth]{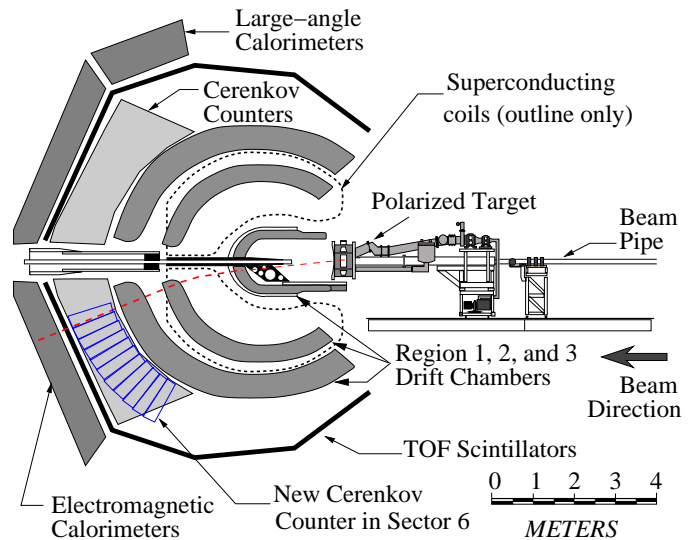}
 \caption{{\it (Color online)} CLAS during EG4 showing the polarized target 
   and the detector arrangement. A new Cherenkov detector consisting of 11 segments
   was installed in place of the original Cherenkov in sector 6. It provided
   the ability of detecting scattered electrons in the outbending configuration
   with scattering angles as small as $6^\circ$ (dashed-line track).
}
 \label{fig:clas}
\end{figure}

The CEBAF Large Acceptance Spectrometer (CLAS) was used to detect scattered
particles~\cite{Mecking:2003zu}. 
Figure~\ref{fig:clas} shows the basic structure of CLAS during EG4 with the
polarized target installed. 
CLAS is an almost hermetic detector, optimized for the measurement of multi-particle
final states in a large momentum region. The detector design is based on a toroidal magnet 
made by six superconducting coils arranged around the beam line to produce a field pointing primarily in the azimuthal direction. The field direction can be set such that the scattered
negatively-charged particles can be either bent away from the beamline (``electron outbending'')
or towards it (``electron inbending''). 
The detector itself is composed of six independent magnetic spectrometers, referred to as 
six ``sectors'', with a common target, trigger, and data acquisition system. Each sector 
is equipped with a three-layer drift chamber (DC) system for momentum and tracking 
determination, a time-of-flight (TOF) counter, a Cherenkov counter (CC) and a 
double-layer electromagnetic calorimeter (EC). The TOF, CC, and EC systems 
are primarily used for determining the particle type. 

\begin{figure}[!ht]
 \includegraphics[width=0.45\textwidth]{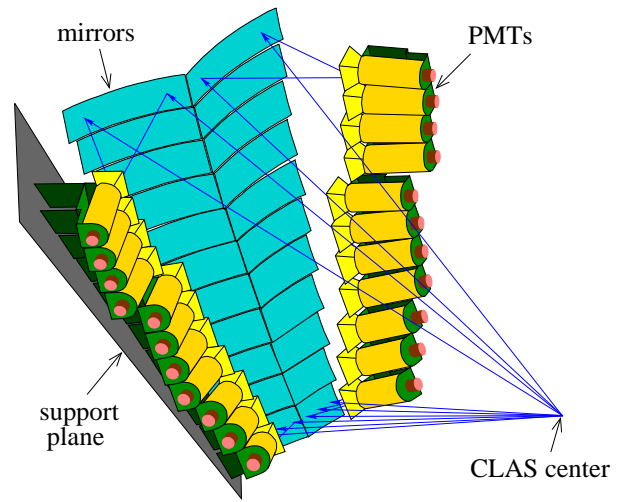}
 \caption{{\it (Color online)} The new Cherenkov detector designed and built by the 
INFN-Genova group. It consists of 11 pairs of mirrors with spherical curvature, which 
reflect the Cherenkov light to corresponding photo-multiplier tubes (PMTs).
Only one of the two support planes for the PMTs is shown here. The solid blue 
lines show simulated particle trajectories originated from the CLAS center and the 
reflection of the Cherenkov light towards the PMT. 
}
 \label{fig:newcer}
\end{figure}

To reach very low $Q^2$ while retaining the high beam energy needed to
measure the GDH sum, a small scattering angle was necessary. This was achieved
by running the CLAS torus magnet in the electron-outbending configuration.
Although the standard CLAS Cherenkov detector geometrically reaches an $8^\circ$ scattering
angle~\cite{Adams:2001kk}, its structure is not ideal for collecting the Cherenkov light
for outbending electrons. Therefore, for the EG4 experiment, a new Cherenkov detector was built
by the INFN-Genova group and installed in sector 6, as shown in Fig.~\ref{fig:clas}.
It was designed to reach $6^\circ$ scattering angle by optimizing the light
collection for the electron-outbending configuration.
Because of the very high counting rates at such low scattering angles, instrumenting only 
one CLAS sector was sufficient for the experiment. 
The new Cherenkov detector used
the same radiator gas (C$_4$F$_{10}$) and the gas flow control system used in the standard
CLAS Cherenkov.
It consisted of 11 segments, each equipped with a pair of light-weight spherical mirrors; 
see Fig.~\ref{fig:newcer}.
The mirrors were constructed following~\cite{Cisbani:2003fj}, by shaping a plexiglass layer
onto a spherical mould, then gluing onto it a sandwich of carbon fiber and honeycomb, and
finally evaporating a thin layer of aluminum onto the plexiglass. Each mirror reflected
the light towards a light collector made of two pieces, an entrance section with the
approximate shape of a truncated pyramid and a guiding section cylindrical in shape
such as to match the circular photocathode. Each light collector was made of plexiglass
with aluminum evaporated on the internal surface. The entrance section was built by a
no-contact technique, where the plexiglass sheet was heated and pushed against a mould
with the desired shape, then the bottom of the obtained object was cut to permit the
free passage of light. The cylindrical section was obtained by cutting a plexiglass tube.
The two sections were then glued together before evaporating the reflective layer.
For the PMTs, the Photonis XP4508B with quartz window were chosen.
The photoelectron yield was greater than $\approx 10$ within the kinematic region of the experiment, 
thereby yielding a high electron detection efficiency down to a scattering angle of about $6^\circ$. 
Signals from the new Cherenkov were built into the main electron trigger during EG4.
Consequently only 1/6 of the full azimuthal acceptance of CLAS was used to detect and
identify forward-angle scattered electrons.

\subsection{The polarized electron beam}
The polarized electron beam was produced by illuminating a strained 
GaAs photocathode with circularly polarized light. The helicity of the
electron beam was selected from a pseudo-random sequence, and followed
a quartet structure of either ``$+--+$'' or ``$-++-$'', with each helicity
state lasting 33~ms. 
The helicity sequence controlled the trigger system, and periods of beam 
instability from helicity reversal were rejected from the data stream. 
To reduce possible systematic uncertainties, data were taken for two
different beam helicity configurations, with the beam insertable half-wave 
plate (IHWP) inserted (in) and removed (out), respectively. 
The polarization of the electron beam was measured by both a M\o ller and a Mott polarimeter. 

\subsection{The polarized targets}\label{sec:target}
\begin{figure}[!htp]
 \begin{center}
 \includegraphics[width=0.45\textwidth]{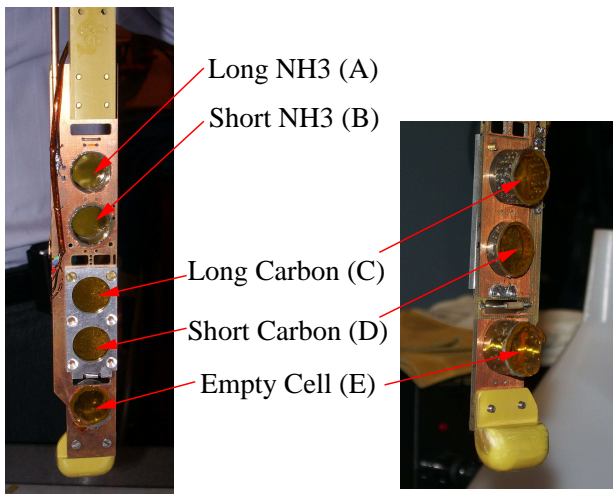}
 \end{center}
 \caption{{\it (Color online)} Target insert used during the EG4 experiment. 
A 1.0-cm long NH$_3$ and the 0.5-cm long NH$_3$ targets were installed 
in the Long and Short NH$_3$ positions during the first half of the NH$_3$ 
run period. They were called the ``long NH$_3$ top'' 
and the ``short NH$_3$'' targets, respectively. During the second half of the 
NH$_3$ run, two 1.0-cm long NH$_3$ targets were installed in the Long and the Short 
positions; they were called the ``long NH$_3$ top'' and the ``long NH$_3$ 
bottom'' 
targets, respectively. For the ND$_3$ run period only one 1.0-cm long ND$_3$ target 
was installed in the Short position. The five target positions are labeled
A, B, C, D, and E, as shown.}\label{fig:targinsert}
\end{figure}

The polarized targets used for EG4 were the frozen $^{15}$NH$_3$ and $^{14}$ND$_3$ targets 
dynamically polarized at 1~K with a 5-Tesla field. These were the same as the 
targets used for previous CLAS double-polarization measurements~\cite{Crabb:1995xi}.
The target material was irradiated with 20~MeV electrons
prior to the experiment to impart the paramagnetic radicals necessary for dynamic polarization. 
It was subsequently stored in liquid nitrogen (LN$_2$) until needed for the experiment.  
The material, in the form of 1-2 mm sized granules, was then removed from the LN$_2$ storage 
dewars and loaded into two cylindrical containers on the target insert. The structure of 
the target insert is shown in Fig.~\ref{fig:targinsert}. The containers were either 1.0 cm 
or 0.5 cm in length, hereafter referred to as the long and short cells, respectively.  The 
insert was then quickly placed into the target ``banjo'', a 1-2 liter vessel of 1-K liquid helium at 
the center of a 5-T superconducting split coil magnet.  A complete description of the 
polarized target can be found in Ref.~\cite{Keith:2003ca}.

Because of the presence of gaps between the frozen crystals inside the target cell, 
even if the length of the 
target cell or the banjo could be determined precisely, the exact amount of polarized 
materials interacting with the electron beam could not be directly measured.
The fraction of the target filled by frozen polarized material is called the 
``packing factor'' and is typically extracted by comparing the yield from
the polarized target to those from 
carbon and ``empty'' targets. For the carbon target, a carbon 
foil with known thickness was placed in an empty target cell and filled with 
liquid $^4$He. There were two carbon targets, labeled ``long'' and ``short'' 
carbon, of which both the cell length and the foil thickness match those of 
the long and the short NH$_3$ targets, respectively. 
Empty targets refer to target cells with no solid material 
inside. Empty targets 
can either be filled with liquid $^4$He, or the $^4$He can be completely
pumped out. There was only one empty cell during EG4 to physically host 
the empty targets, which was 1.0 cm in length. 

During EG4 the polarized target was placed 1.01~m upstream from the CLAS 
center to increase the acceptance at low $Q^2$ by reducing the minimum angle 
for the scattered electrons. The following targets were used: two 1.0-cm long
and one 0.5-cm long NH$_3$ target, one 1.0-cm long ND$_3$ target, one $0.216$-cm and 
one 0.108-cm thick $^{12}$C target, and one empty target.
The target types during EG4 are defined in Table~\ref{tab:target_type}. 
Unless specified otherwise, ``empty target'' refers to target type 3 
[empty cell with helium (1~cm)] hereafter.
\begin{table}
\begin{center}
 \caption{Targets used during EG4 along with their target lengths and densities. 
The target ID was the value recorded in the data. ID 10 was not used. 
The target position refers to the physical location on the target 
insert defined in Fig.~\ref{fig:targinsert}.}\label{tab:target_type}
 \begin{tabular}{|c|c|c|c|c|}\hline \hline
   Target    &  Target type      & Target & length  & Density \\
     ID      &                   & position & (cm)    & (g/cm$^3$) \\ \hline
     1       & long NH$_3$ top        & A & 1.0 & 0.917$^a$   \\ \hline
     2       & long ND$_3$            & B & 1.0 & 1.056$^a$    \\ \hline
     3       & empty cell with helium  & E & 1.0 & 0.145$^b$ \\ \hline
     4       & long carbon            & C & 1.0, 0.216$^c$ & 2.166$^d$ \\ \hline
     5       & short NH$_3$            & B & 0.5  & 0.917$^a$ \\ \hline
     6       & short carbon           & D & 0.5, 0.108$^c$ & 2.166$^d$ \\ \hline
     7       & long carbon no helium  & C & 1.0, 0.216$^c$ & 2.166$^d$ \\ \hline
     8       & empty cell without helium & E & 1.0 & \\  \hline
     9       & short carbon without helium & D & 0.5 & 2.166$^c$ \\ \hline
     11      & long NH$_3$ bottom & B & 1.0 & 0.917$^a$    \\\hline \hline
 \end{tabular}
\end{center}
\begin{flushleft}
$^a$ For polarized NH$_3$ or ND$_3$ the densities are the density of the frozen polarized 
material beads.\\
$^b$ Helium density.\\
$^c$ The first and the second length values correspond to the cell length and the
carbon foil thickness, respectively.\\
$^d$ Carbon density.
\end{flushleft}
\end{table}

An NMR system was used to monitor the polarization of the target during the experiment, but was 
subject to three systematic uncertainties that limited its suitability for data analysis.  
First, the NMR coils were wrapped around the outside of the 1.5-cm diameter target cells, 
while the electron beam was only rastered over the central 1.2 cm portion of the target.  
The NMR signal was thus dominated by the material at the edges of the cell, and lacked 
sensitivity to the beam-induced depolarization of the material at the center.  This 
uncertainty is difficult to estimate, as the effect depends on the accumulated dose. 
Second, for the EG4 experiment the two polarized target cells were adjacent to one another 
on the insert, as shown in Fig.~\ref{fig:targinsert}, and cross-talk was observed between 
the cells' NMR circuits.  Tests performed at the end of the experiment indicate that
cross-talk could contribute an uncertainty of about 5-10\% to the polarization measurement 
because of 
its effect on the thermal-equilibrium calibration of the NMR signal.  Third, calibration of 
the NMR system itself is normally subject to a 4-5\% uncertainty.  
These three effects added up to a large systematic 
uncertainty to the target polarization measured by NMR. 
Therefore, it was 
decided that the asymmetries of $ep$ elastic scattering would be used to extract the product 
of the beam and target polarizations $P_bP_t$ needed for the exclusive channel 
analysis reported here. 
The methods and results for the elastic $P_bP_t$ extraction will be described in 
Sec.~\ref{sec:ana_pbpt}. 
For NH$_3$, the use of $^{15}$N has the advantage that only one unpaired proton can 
be polarized, while all neutrons are paired to spin zero.  The polarized proton 
in the $^{15}$N does, however, affect the measured asymmetry by a small amount, as 
discussed in Sec.~\ref{sec:ana_15N}.

\section{Data Analysis}
\subsection{Exclusive event selection}
Exclusive events $\vec e\vec p\to e^\prime\pi^+(n)$ were identified 
by detecting the final state electron in coincidence with a pion and using
a missing mass cut to select the undetected neutron. 
For each event, we required that two particles be detected with 
the correct charges ($-1$ for the electron and $+1$
for the $\pi^+$). Each particle was required to have valid information from 
DC and TOF, and have reconstructed momentum greater than 0.3~GeV/$c$ 
(0.1~GeV/$c$ higher than the momentum acceptance of CLAS~\cite{Mecking:2003zu}). 

For particle identification, EC and CC signals were used to identify electrons. 
Cuts were applied on the EC: $E_{tot}>(p-0.3)\times 0.22$, 
$E_{in} > (0.14 p-0.8 E_{out})$ and $E_{in} > 0.035 p$, where 
$E_{in}$ and $E_{out}$ are the energy deposited in the 
inner and the outer layers of the EC, respectively; $E_{tot}=E_{in}+E_{out}$ and 
$p$ is the particle momentum in GeV/$c$. These cuts were selected to optimize
the separation of electrons (that produce electromagnetic showers) from 
pions (that deposit energy mostly through ionizations). 
We also required there to be 
only one hit in the CC, with its signal consistent 
with those from the EC and the TOF in both hit position and timing. 

\begin{figure}[!ht]
 \includegraphics[width=0.45\textwidth]{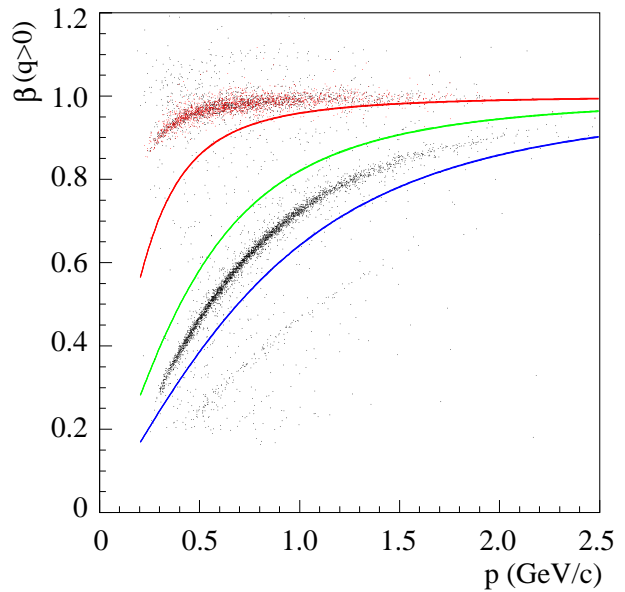}
 \caption{$\beta$ vs. $p$ for all positively charged particles,
with (red) and without (black) TOF cut for pions.
The red, green and blue curves correspond to reconstructed masses
of $0.3$, $0.7$ and $1.2$~GeV$/c^2$, respectively, which are typical cut-off
values used to distinguish between pions and kaons, kaons and protons, and 
protons and heavier particles. As can be seen, the positively charged 
particles detected consist of significant fractions of protons and heavier 
particles and a small fraction of kaons, but the $\pm 1.0$~ns TOF cut is
quite effective in selecting pions. These data were collected on the 
long top NH$_3$ target during the 3~GeV run period.
}\label{fig:beta_vs_p_tof}
\end{figure}

Pions were determined from a mass cut of $0.01<m<0.30$~GeV/$c^2$ and 
a TOF cut $|t_{TOF}-t_{\mathrm{expected}}^\pi|<1.0$~ns. The expected 
flight time of the pion, $t_{\mathrm{expected}}^\pi$, was calculated from the 
particle's momentum in combination with the timing of the electron. 
Figure~\ref{fig:beta_vs_p_tof} shows the effect of the TOF cut
 on the $\beta\equiv v/c$ vs. momentum $p$ distributions, where 
$v$ is the velocity amplitude (speed) of the particle. The TOF cut used
clearly selected pions out of other particle background.

For each event, a vertex $z$ was used. Here $z$ is defined as pointing 
along the beam direction with the origin coinciding with the CLAS center.  
The polarized target was positioned upstream of the CLAS center during EG4 
(see Fig.~\ref{fig:clas}), and the center of the target was determined from 
empty target data to be at $z=-101$~cm. The $z$ cut was optimized to be 
\begin{eqnarray}
 -106~\mathrm{cm}<z<-96~\mathrm{cm}~, \label{eq:vzcut}
\end{eqnarray}
where the range was determined using empty target data to exclude as much 
material outside the target as possible.  See Fig.~\ref{fig:raster_corr} 
in Sec.~\ref{sec:kine_corr} for a detailed presentation of the vertex $z$
distribution. 

Acceptance cuts, also called ``fiducial cuts'', were applied on both 
electrons and pions using reconstructed
DC variables. These acceptance cuts exclude regions where the detector 
efficiency is not well understood, which often happens on the edge 
of the detectors, but could also include regions
where certain parts of the detectors malfunctioned. 
Moreover, because the main purpose of EG4 was measurement of the GDH sum, which 
only requires detection of inclusively-scattered electrons, not all six  
DC sectors were turned on during the run. This caused a variation in the 
$\phi^*$ acceptance of the exclusive channel.
Determination of the acceptance and its effects on the asymmetries will be described in 
Sec.~\ref{sec:accp_corr}.

\subsection{Beam properties}

As described in the previous section, the helicity of the electron beam
followed a quartet structure. For EG4, the beam helicity of each event 
was delayed by 8 pulses (2 quartets) and then recorded in the data stream. 
This delayed recording helped to avoid cross-talk between the helicity 
signal and the electronics or data acquisition system in the hall.
In the data analysis, the delay of the helicity sequence was corrected 
to match each event to its true beam helicity state. During this process, 
events with inconsistent recording of the helicity sequence were rejected. 

A helicity dependence of the integrated beam charge causes a first-order correction to the measured
physics asymmetry, and thus it is desired to keep the charge asymmetry 
as small as possible. The beam charge asymmetry was calculated using the 
charge measured by the Faraday cup. It was found to be below the percent
level throughout the EG4 experiment, and for most runs had stable values 
at or below the $10^{-3}$ level. 

\begin{figure}[!htp]
 \begin{center}
 \includegraphics[width=0.45\textwidth]{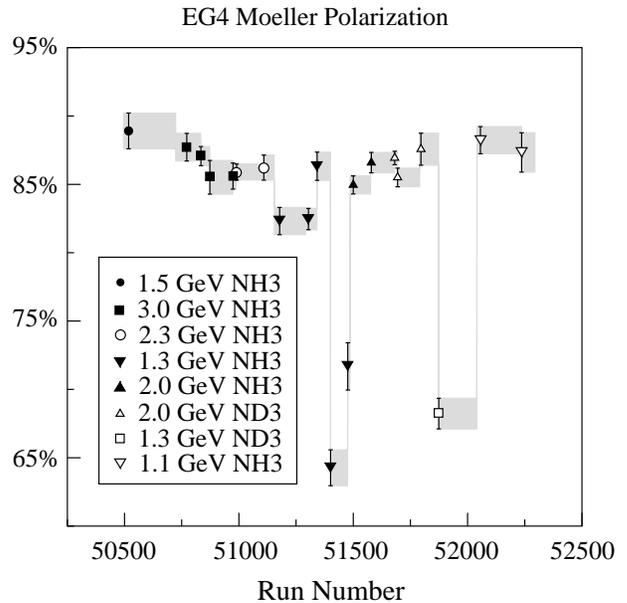} 
 \end{center}
 \caption{Beam polarization from M\o ller measurements vs. run number for the whole EG4 experiment. 
   The gray bands represent extrapolations of the beam polarization to the corresponding range of runs
   as described in the text.}\label{fig:moller}
\end{figure}

Different methods for deriving the beam energy were 
used during EG4. The exact energies were 1.054, 1.338, 1.989, 2.260 and 2.999~GeV. 
The beam polarization was determined using a M\o ller 
polarimeter~\cite{Mecking:2003zu} in Hall B that measured the asymmetry in 
elastic electron-electron scattering.  
The results are shown in Fig.~\ref{fig:moller}. Typically, M\o ller measurements
were performed as soon as a change to the beam configuration was made, and then
intermittently throughout the run period. Therefore, the beam polarization from
each M\o ller measurement was applied retroactively to runs that immediately follow such configuration
changes, and to runs that follow the M\o ller measurement until
the next valid measurement is available. 
Two additional measurements were done using a Mott 
polarimeter~\cite{Price:1996up,Price:1997qf,Price:1998xd,Steigerwald-MottJLab}, 
which is located near the injector where the beam electrons have reached 5~MeV
in energy but before entering the first linac. The Mott polarimeter results
were consistent with those from M\o ller measurements. 
The absolute beam helicity was determined using the $\sin\phi^*$-weighted
moment of the beam asymmetry
$A_{LU}$ in the $\Delta(1232)3/2^+$ region and comparing with results from previous
experiments~\cite{Joo:2004mi,Park:2007tn}.
%
Using the $A_{LU}$ method, it was determined that when the beam IHWP is
inserted, for beam energies 
1.3 and 2.3 GeV, the positive DAQ helicity corresponds to the
true negative helicity of the beam electron, while for other
energies the postive DAQ helicity corresponds 
to the true positive electron helicity. These results are consistent
with the sign change of the beam polarization measured with the
M\o ller polarimeter.

%

\subsection{Kinematic corrections}\label{sec:kine_corr}
\begin{figure}[!htp]
 \begin{center}
  \includegraphics[width=0.45\textwidth]{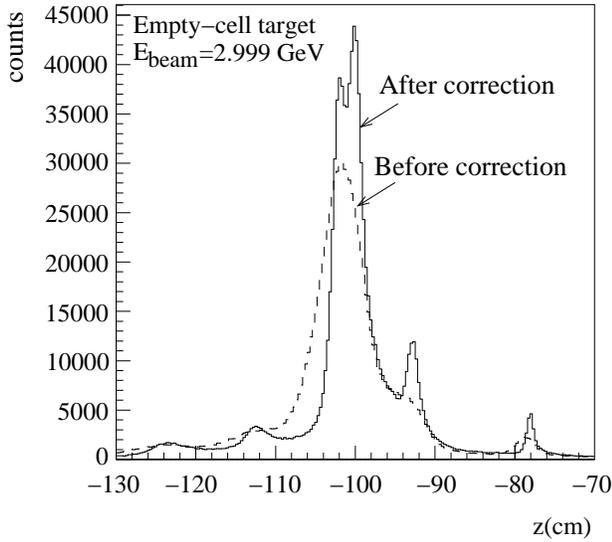}
 \end{center}
  \caption{Electrons' vertex $z$ position before (dashed) and after (solid) 
raster corrections, taken with the empty target with the 3 GeV beam. While the 
beam line exit window (at $z=-78.3$~cm) can be seen both before and after the 
correction, the banjo windows (at $z=-100$ and $-102$~cm), the 4~K heat shield 
(14~$\mu$m aluminum at $z=-121.0$~cm), some target structure at $z\approx -112$~cm, 
and several insulating foils (aluminum or aluminumized mylar, between $z=-90.5$ 
and $-94.1$~cm), become visible only after the raster correction. The vertex $z$ 
cut, Eq.~(\ref{eq:vzcut}), corresponds to slightly more than $3\sigma$ in the 
target thickness~\cite{thesis:adhikari}.}\label{fig:raster_corr}
\end{figure}

Various corrections were applied to the kinematic variables 
reconstructed from the detectors~\cite{thesis:adhikari}. 
The first is the raster correction: To avoid the electron beam 
overheating the target, the beam was rastered in a circular pattern during 
EG4 using four magnets located upstream of the target.  The values of the 
magnet current were recorded in the data stream and were used to calculate 
the beam position $(x,y)$ at the target. The beam position was then used to 
re-calculate the vertex position along the beam direction $z$. 
After the raster correction was applied, the average value of 
the $z$ positions of all particles in the same event was taken 
as the true vertex position of the event, 
see Fig.~\ref{fig:raster_corr}~\cite{thesis:adhikari}. The polar and the azimuthal 
angles $\theta$ and $\phi$ 
of each particle were also corrected using the new beam and vertex positions. 
This procedure took into account the multiple scattering effect 
that affected the reconstructed vertex position randomly for each particle. 

Because of uncertainties in our knowledge of the drift chamber positions and of the
shape and location of the torus coils, a systematic shift of the particle 
momentum was present. To correct for this shift, 
the magnitude of the reconstructed particle momentum $p$ and the polar angle 
$\theta$ were adjusted using sector-dependent parameters. The detailed method
for the momentum correction is described in Ref.~\cite{CLASNOTE03-005} and results
for this experiment are given in Ref.~\cite{thesis:adhikari}. 
For sector 6 equipped with the new Cherenkov counter, 
inclusive elastic $ep$ scattering events were used to optimize the 
correction based on the invariant mass $W$ position of the elastic 
peak. For the other sectors, electron triggers were not available and 
hadrons from exclusive events such as $ep\to e'p'X$, 
$ep\to e'\pi^+\pi^-X$, and exclusive events
$ep\to e'p' \pi^+ \pi^-$ were used to optimize the corrections. 

Finally, the momentum of each particle was corrected for the energy loss 
from passage through material enclosed in the target banjo and the target windows. 
For electrons a single value ${dE}/{dx}= 2.8$~MeV/(g/cm$^2$) 
was used, while for other particles the Bethe-Bloch equation~\cite{Beringer:2012} 
was used to calculate the ionization loss. 

\begin{figure}[!htp]
 \begin{center}
 \includegraphics[width=0.45\textwidth]{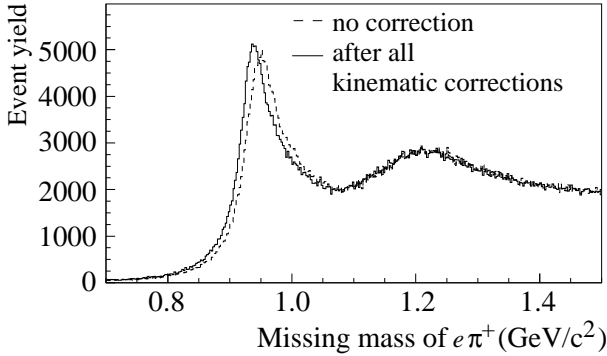}
 \end{center}
  \caption{Missing mass spectrum for the $e+p\to e^\prime \pi^+(X)$ channel 
before (dashed) and after all kinematics corrections (solid), from six 
3.0-GeV long top NH$_3$ target runs. After all corrections, the peak center
is closer to the expected value (the neutron mass).}\label{fig:mx_pcorr}
\end{figure}

Figure~\ref{fig:mx_pcorr} shows the effect on the missing mass spectrum 
for the $ep\to e^\prime \pi^+(X)$ channel from kinematic corrections.

\subsection{Elastic scattering for extracting $P_bP_t$}\label{sec:ana_pbpt}
The product of the beam and the target polarizations $P_bP_t$ is needed to directly 
correct the exclusive channel asymmetries. 
During EG4, the target polarization $P_t$ was measured by NMR and the beam polarization $P_b$ 
by the M\o ller polarimetry. However, due to reasons described in Sec.~\ref{sec:target}, 
the NMR measurements had large uncertainties and an alternate method had to be used.
For EG4 we extracted $P_bP_t$ for all beam energies by comparing the double spin
asymmetry of elastic $ep$ events to the expected value: 
\begin{eqnarray}
   P_bP_t &=& \frac{A^{el}_{\mathrm{meas}}}{A^{el}_{\mathrm{th}}}~,
\end{eqnarray}
where the measured elastic asymmetry was extracted from data using
\begin{eqnarray}
  A^{el}_{\mathrm{meas}}&=&\frac{A^{el}_{\mathrm{raw}}}{f_{el}}, \label{eq:Ael_dil}
\end{eqnarray}
with $f_{el}$ the elastic dilution factor to account for the effect of events scattered 
from unpolarized material in the target. 
The raw asymmetry was evaluated as 
\begin{eqnarray}
   {A^{el}_{\mathrm{raw}}} &=& 
     \frac{\frac{N_R^{el}}{Q_R}-\frac{N_L^{el}}{Q_L}}
          {\frac{N_R^{el}}{Q_R}+\frac{N_L^{el}}{Q_L}}~,
\end{eqnarray}
where $N_{R(L)}^{el}$ and $Q_{R(L)}$ are the elastic event yield and the 
beam charge for the right- (left-)handed beam electrons, respectively. 
The expected elastic-scattering asymmetry $A^{el}_{\mathrm{th}}$ 
was calculated using
\begin{eqnarray}
  &A^{el}_{\mathrm{th}}&= -2\sqrt{\frac{\tau}{1+\tau}}\tan\frac{\theta_e}{2}\nonumber\\
  &\times&{\left[\sqrt{\tau\left(1+(1+\tau)\tan^2\frac{\theta_e}{2}\right)}\cos\theta_e 
  +\sin\theta_e {{G_E^p}\over{G_M^p}}\right]}\over
  {\left[\frac{\left(G_E^p/G_M^p\right)^2+\tau}{1+\tau}+2\tau \tan^2\frac{\theta_e}{2}\right]}~,
\end{eqnarray}
with $\tau=Q^2/(4M^2)$.
The proton form factor fits from Ref.~\cite{Bosted:1994tm} were used:  
\begin{eqnarray}
  G_E^p &=& \left.{1}\middle/ \left[1+0.62Q+0.68 Q^2+2.8 Q^3 + 0.83 Q^4\right]\right.
\end{eqnarray}
and
\begin{eqnarray}
  G_M^p &=& \left.{2.79}\middle/\left[1+0.35Q+2.44Q^2\right.\right.\nonumber\\
  &~& \left.+0.5Q^3+1.04Q^4+0.34Q^5\right]
\end{eqnarray}
with $Q\equiv \sqrt{Q^2}$ in GeV/$c$. 
Using a more updated fit of the proton form factors than Ref.~\cite{Bosted:1994tm} 
would change the asymmetry value by less than 2\% relative.

Elastic events were identified using two methods: (1) inclusive elastic events where only the 
scattered electron was detected and a cut on the invariant mass $W$ near the proton peak 
was applied;
(2) exclusive elastic events where both the scattered proton and electron were detected and
cuts were applied to the electron and the proton azimuthal angles: $\vert\phi_e-\phi_p-180^\circ\vert<3^\circ$,
the polar angles of the proton and the electron's momentum transfer $\vec q$, 
$\vert\theta_p-\theta_q\vert<2^\circ$, and the missing energy $E_\mathrm{miss}<0.15$~GeV. The exclusive 
analysis had limited statistics and only worked for the 3.0 and the 2.3 GeV data sets. For
lower beam energies, the proton's scattering angle was typically greater than $49^\circ$, 
and was blocked by the polarized target coils. Therefore the $P_bP_t$ value extracted from exclusive
elastic events was only used as a cross-check of the $P_bP_t$ from inclusive events. 

The presence of unpolarized material reduces the measured asymmetry, and this effect is
described as a dilution factor in the analysis. 
The dilution factor for the inclusive elastic events, $f_{el}^\mathrm{incl}$, was extracted by 
comparing the invariant mass $W$ spectrum of the polarized target to that computed for
the unpolarized material. 
The beam-charge-normalized $W$ spectrum for the unpolarized material in the polarized 
target, denoted as $\frac{N_{\mathrm{N~in~\mathrm{NH}}_3}}{Q_{\mathrm{NH}_3}}$, 
was calculated using the spectra of the carbon and the empty target, 
the known thickness and density of the carbon and the empty
target, and the polarized target's packing factor $x_\mathrm{NH_3}$ 
defined as the absolute length of the polarized material in the polarized target:
\begin{eqnarray}
  \frac{N_{\mathrm{N~in~\mathrm{NH}}_3}}{Q_{\mathrm{NH}_3}} & = &
     r_C \frac{N_{^{12}\mathrm{C}}}{Q_{^{12}\mathrm{C}}}
    + r_\mathrm{empt} \frac{N_\mathrm{empt}}{Q_\mathrm{empt}}~,\label{eq:eldil_incl}
\end{eqnarray}
where $N_{^{12}\mathrm{C}(\mathrm{empt})}$ and ${Q_{^{12}\mathrm{C}(\mathrm{empt})}}$ are
the yield and the beam charge of the carbon (empty) target data. The scaling factors are
\begin{eqnarray}
 r_{C}&=&\frac{\left({B_{\mathrm{NH}_3}}{\rho_{\mathrm{NH}_3} x_{\mathrm{NH}_3}}
	 + {{B_w}\rho_w x_w}  \frac{x_{\mathrm{NH}_3}}{l}\right)}
         {{B_{^{12}\mathrm{C}}}\rho_{^{12}\mathrm{C}} x_{^{12}\mathrm{C}}
          + {{B_w}\rho_w x_w}\frac{x_{^{12}\mathrm{C}}}{l}}~,
         \label{eq:eldil_rC}\\
r_\mathrm{empt} &=& \left(1-\frac{x_{\mathrm{NH}_3}}{l}\right)-\left(1-\frac{x_{^{12}\mathrm{C}}}{l}\right)r_C
~,\label{eq:eldil_rEmpt}
\end{eqnarray}
where $x_{^{12}\mathrm{C}}$ 
is the thickness of the carbon foil in the carbon target, $x_w$ is the sum of thicknesses of
other unpolarized material in the target, $l$ is the target 
banjo length (1.0~cm for the long target and 0.5~cm for the short target), 
and $B_{^{12}\mathrm{C},w}=1$ are the bound-nucleon fractions of the 
carbon target and other unpolarized material in the target, respectively. 
The values 
of $x$ for the various materials are given in Table~\ref{tab:material}.
The bound-nucleon fraction for the NH$_3$ target takes into account both 
the fraction of bound nucleons and a correction for the extra 
neutron in the $^{15}$N: $B_{\mathrm{NH}_3}=(14+\sigma_n/\sigma_N)/18$ with $\sigma_N=(\sigma_p+\sigma_n)/2$ 
and $\sigma_{p,n}$ are the calculated elastic cross sections for the proton and the neutron, respectively. 

\begin{table}[!htp]
\caption{Material used for the EG4 target and their locations in increasing order of $z$, 
in the range $z=(-120,-80)$~cm. The ratios $Z/A$ were used in the dilution factor analysis of the 
exclusive channel, see Sec.\ref{sec:ana_dilution}.
}\label{tab:material}
\begin{center}
\begin{tabular}{p{1cm}|p{2cm}|p{1.2cm}|p{2cm}|p{1.6cm} } \hline
 location $z$ (cm)& Material & Density (g/cm$^3$) & Thickness         & $Z/A$ \\ \hline
 -101.9 & banjo entrance window, Al        & $2.7$   & $71$ $\mu$m        & $13./26.982$ \\
 varies  & target entrance window, kapton  & $1.42$  & $25$ $\mu$m        & $0.51264$  \\
 varies  & NH$_3$                         & $0.917$ & $x^a$ & $7/18$ \\
 varies  & long $^{12}$C                   & $2.166$ & $2.16\pm 0.05$~mm & $6/12$ \\
 varies  & liquid $^4$He                  & $0.145$ & $l-x^a$           & $2/4$ \\
 varies  & target entrance window kapton  & $1.42$  & $25$ $\mu$m        & $0.51264$ \\
 -99.6   &  banjo exit window Al          & $2.7$   & $71$ $\mu$m        & $13./26.982$ \\
 \hline
\end{tabular}
\end{center}
$^a$ $l$ is the banjo length and $x$ is either the packing factor (for NH$_3$ targets) or the carbon 
foil thickness (for carbon targets).
\end{table}

After the contribution from the unpolarized material was known, the dilution factor was 
calculated using 
\begin{eqnarray}
  f_{el}^\mathrm{incl} &=& \frac{N_\mathrm{p~in~NH_3}}{N_{\mathrm{NH}_3}}
   = \frac{N_{\mathrm{NH}_3}-N_{\mathrm{N~in~NH}_3}}{N_{\mathrm{NH}_3}}~,
\end{eqnarray}
where $N_{\mathrm{NH}_3}$ is the total number of events from the NH$_3$ target. 
The dilution correction to the elastic asymmetry was then applied using Eq.~(\ref{eq:Ael_dil}). 
In the present analysis, elastic events below $Q^2=0.156$~(GeV/$c$)$^2$ could not be used because of 
electrons scattered elastically from nuclei in the target, such as $^4$He and nitrogen. 
These low $Q^2$ bins were rejected in the $P_bP_t$ analysis.

\begin{figure}[!htp]
\includegraphics[width=0.45\textwidth]{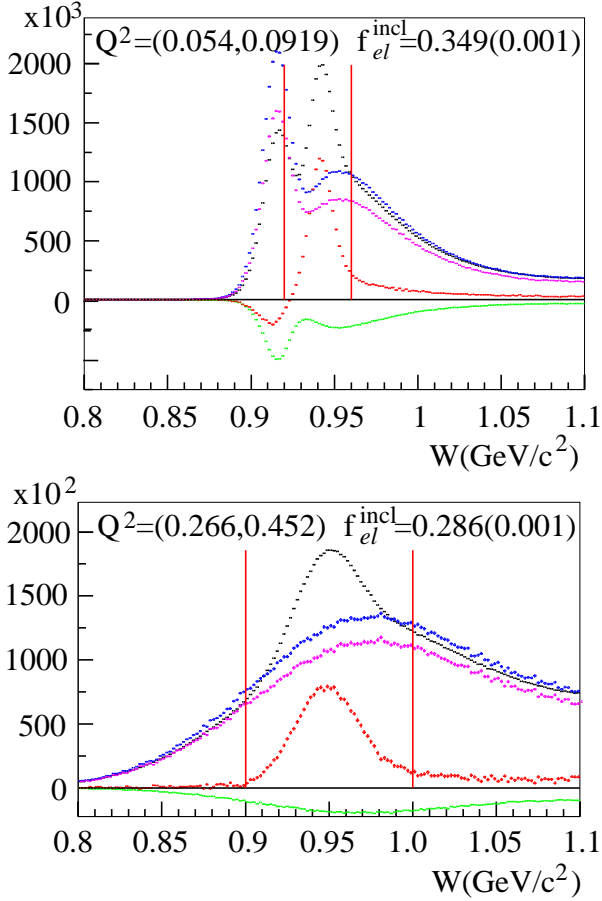}
 \caption{$W$ spectrum for dilution calculation for inclusive elastic $P_bP_t$ analysis.
(Top) 1.1-GeV data on NH$_3$ long bottom target in the $Q^2=(0.054,0.092)$~(GeV/$c$)$^2$ bin; 
(bottom) 3.0-GeV data on NH$_3$ long top target in the $Q^2=(0.266,0.452)$~(GeV/$c$)$^2$ bin.
For each panel, histograms from the carbon target (blue) and empty target 
(green) were scaled using Eqs.~(\ref{eq:eldil_rC}) and (\ref{eq:eldil_rEmpt}) 
using a packing factor of 0.75~cm for 1.1 GeV and 
0.65~cm for 3.0 GeV, respectively, and their sum gave the estimated contribution
from unpolarized material in the NH$_3$ target (magenta). This unpolarized background 
was then subtracted from the NH$_3$ spectrum (black) to estimate the contribution 
from polarized protons in the target (red).
The calculated elastic dilution factors are shown for each set of data with their uncertainties
in the brackets. 
The $W$ cuts
used to select elastic events are shown as the two red vertical lines. 
Note that the scaled empty target spectrum (green) is negative, indicating that for the 
chosen packing factor we have scaled up the carbon data and then subtracted the extra helium
to reproduce the unpolarized background in NH$_3$. For $Q^2$ bins below 
0.156~(GeV/$c$)$^2$, the nuclear elastic event contaminates the $ep$ elastic peak 
and the extraction of the dilution
factor is not reliable. For this reason, data with $Q^2<0.156$~(GeV/$c$)$^2$ 
were rejected from the elastic $P_bP_t$ analysis.
}\label{fig:elw_incl}
\end{figure}

\begin{figure}[!htp]
  \includegraphics[width=0.45\textwidth]{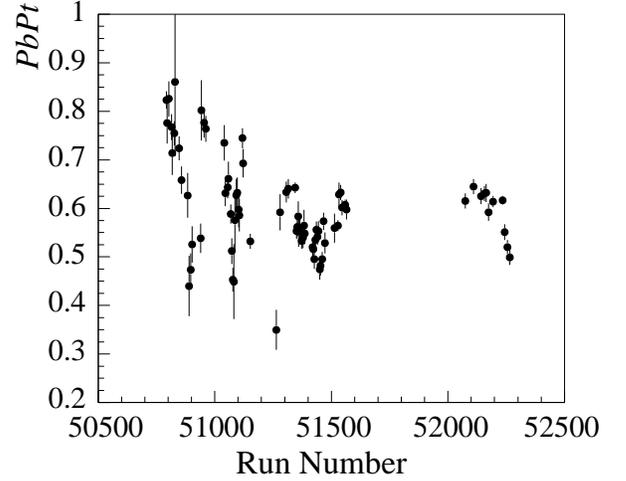}
  \caption{Magnitude of $P_bP_t$ extracted from inclusive elastic scattering events for
    all runs used in the present analysis that were taken on 
    the polarized NH$_3$ target. 
    For illustration purposes, results from adjacent
    runs that shared the same beam insertable half-wave plate status were combined and
    are shown as one data point here. 
    The error bars shown are statistical uncertainties
    determined by the number of available elastic events.}\label{fig:pbpt_allruns}
\end{figure}

Figure~\ref{fig:elw_incl} shows the $W$ spectrum decomposition for 1.1 and 
3.0~GeV inclusive elastic scattering data for two $Q^2$ bins. The low
$Q^2$ bin (top) is to illustrate the effect of the nuclear elastic scattering and these bins
were rejected from the $P_bP_t$ analysis. The high $Q^2$ bin (bottom) shows no such effect 
and the $P_bP_t$ extracted are considered reliable. 
After the $P_bP_t$ value was extracted for individual $Q^2$ bins, the results were checked
to ensure there was no systematic $Q^2$-dependence, which would imply a problem with the 
analysis. 
The $P_bP_t$ results were then averaged over all $Q^2$ bins above $0.156$~(GeV/$c$)$^2$. 
This was done for each individual run and the run-by-run, 
$Q^2$-averaged $P_bP_t$ results were used to correct the asymmetries from 
the exclusive channel.
Figure~\ref{fig:pbpt_allruns} illustrates the variation of $P_bP_t$ during the experiment.

The uncertainty of the packing factor $x_\mathrm{NH_3}$ used in the analysis was checked 
using the $W$ spectrum below $W=0.9$~(GeV/$c^2$), because an incorrect normalization 
would yield an over- or
an under-subtraction of the yield from unpolarized material.
For the 2.3 and 3.0~GeV data the value of $x_\mathrm{NH_3}$ was confirmed by 
comparing the $P_bP_t$ value extracted from the inclusive to that from the exclusive elastic events. 
The packing factor and its uncertainty also affect the dilution analysis
of the exclusive channel, to be described in the next sections, thus the 
final results on $P_bP_t$ for each combination 
of beam energy and polarized target type are shown together
with the exclusive channel dilution results in Table~\ref{tab:PbPtfdil}. 
The relatively larger error bar for the 1.1 GeV NH$_3$ long bottom target is 
because most of the data were affected by the nuclear elastic scattering and 
there are very limited $Q^2$ bins available for the elastic $P_bP_t$ analysis.

In addition to checking the $W$ spectrum and the comparison between inclusive and exclusive
elastic events, the $en\to e^\prime\pi^-(p)$ channel was also used to check $x_\mathrm{NH_3}$ because
these events come primarily from the unpolarized neutrons of the nitrogen 
in the target and thus should have a dilution factor of zero. 
The $e^\prime\pi^-(p)$ events were analyzed for all beam energies 
and it was found the dilution factors calculated using the $x_\mathrm{NH_3}$ values
in Table~\ref{tab:PbPtfdil} were indeed consistent with zero.
As a last check, the run-by-run values of $P_bP_t$ were compared with the 
numerous target material and configuration changes during the 
experiment, and were found to be consistent with the physical changes of the target. 

\subsection{Extraction of exclusive channel asymmetries}\label{sec:ana_procedure}
To extract the exclusive channel asymmetries, the $e^\prime \pi^+ (n)$ channel events
were divided into four-dimensional bins in $W$, $Q^2$, $\cos\theta^*$, and $\phi^*$
and then the asymmetries were extracted from the counts in each bin. 
The event counts for the four combinations of
beam helicities and target polarization can be written, based on Eq.~(\ref{eq:sigma_h}), as
\begin{eqnarray}
 N_{\ua\Ua} &=& D_1 \left[\sigma_0 + P_b^\Ua \sigma_e + f_\mathrm{dil}^\pi P_t^\Ua \sigma_t + P_b^\Ua f_\mathrm{dil}^\pi P_t^\Ua \sigma_{et} \right]~, \label{eq:Nuu}\\
 N_{\da\Ua} &=& D_2 \left[\sigma_0 - P_b^\Ua \sigma_e + f_\mathrm{dil}^\pi P_t^\Ua \sigma_t - P_b^\Ua f_\mathrm{dil}^\pi P_t^\Ua \sigma_{et} \right]~, \label{eq:Ndu}\\
 N_{\ua\Da} &=& D_3 \left[\sigma_0 + P_b^\Da \sigma_e - f_\mathrm{dil}^\pi P_t^\Da \sigma_t - P_b^\Da f_\mathrm{dil}^\pi P_t^\Da \sigma_{et} \right]~, \label{eq:Nud}\\
 N_{\da\Da} &=& D_4 \left[\sigma_0 - P_b^\Da \sigma_e - f_\mathrm{dil}^\pi P_t^\Da \sigma_t + P_b^\Da f_\mathrm{dil}^\pi P_t^\Da \sigma_{et} \right]~, \label{eq:Ndd}
\end{eqnarray}
where the arrows in the subscripts of $N$ are for the beam helicities 
($\ua$ or $\da$) and the target spin directions ($\Ua$ or $\Da$), respectively, 
with $\ua$ and $\Ua$ being positive helicity or parallel to the beam direction 
and $\da$ and $\Da$ being negative helicity or anti-parallel to the beam direction. 
The parameters $P^\Ua$ and $P^\Da$ are the statistically-averaged target or beam 
polarizations when the target spin is aligned and anti-aligned to the beamline, 
respectively. The dilution factor $f_\mathrm{dil}^\pi$ for the exclusive channel 
$\vec e\vec p\to e^\prime \pi^+(n)$ is defined
as the fractional yield from the polarized proton in the NH$_3$ target, which 
effectively changes the target polarization. 
The four parameters $D_{1,2,3,4}$, relating event counts to cross sections, are 
related to
the total beam charge, target thickness, spectrometer acceptance, and detector 
efficiencies for each configuration. For stable running periods with no significant 
change in the target cell, the spectrometer setting and the detector status, the $D$ 
factor is strictly proportional to the accumulated beam charge in each setting. 
%
%
From Eqs.~(\ref{eq:Nuu})--(\ref{eq:Ndd}), one can form the asymmetries as:
\begin{eqnarray}
  A_{LU} &=& \frac{1}{P_b^\Ua P_b^\Da}\times\nonumber\\
   &&\hspace*{-1cm}
   \left[\frac{\left(\frac{N_{\ua\Da}}{D_3}-\frac{N_{\da\Da}}{D_4}\right) P_b^\Ua P_t^\Ua + 
   \left(\frac{N_{\ua\Ua}}{D_1}-\frac{N_{\da\Ua}}{D_2}\right) P_b^\Da P_t^\Da}
   {\left(\frac{N_{\ua\Ua}}{D_1}+\frac{N_{\da\Ua}}{D_2}\right) P_t^\Da +
   \left(\frac{N_{\ua\Da}}{D_3}+\frac{N_{\da\Da}}{D_4}\right) P_t^\Ua}\right]~,
    \label{eq:ae}\\
  A_{UL} &=& \frac{1}{f_\mathrm{dil}^\pi}\frac{\left(\frac{N_{\ua\Ua}}{D_1} + \frac{N_{\da\Ua}}{D_2}\right)
      -\left(\frac{N_{\ua\Da}}{D_3} + \frac{N_{\da\Da}}{D_4}\right)}
   {\left(\frac{N_{\ua\Ua}}{D_1}+\frac{N_{\da\Ua}}{D_2}\right) P_t^\Da +
   \left(\frac{N_{\ua\Da}}{D_3}+\frac{N_{\da\Da}}{D_4}\right) P_t^\Ua}~,
    \label{eq:at}\\
  A_{LL} &=& \frac{1}{P_b^\Ua P_b^\Da f_\mathrm{dil}^\pi}\times\nonumber\\
   && \left[
   \frac{\left(\frac{N_{\ua\Da}}{D_3}-\frac{N_{\da\Da}}{D_4}\right) P_b^\Ua - 
   \left(\frac{N_{\ua\Ua}}{D_1}-\frac{N_{\da\Ua}}{D_2}\right) P_b^\Da}
   {\left(\frac{N_{\ua\Ua}}{D_1}+\frac{N_{\da\Ua}}{D_2}\right) P_t^\Da +
   \left(\frac{N_{\ua\Da}}{D_3}+\frac{N_{\da\Da}}{D_4}\right) P_t^\Ua}
   \right]~.
   \label{eq:aet}
\end{eqnarray}

\subsection{Dilution factor for the exclusive channel}\label{sec:ana_dilution}
\begin{figure}[!htp]
 \begin{center}
  \includegraphics[width=0.45\textwidth]{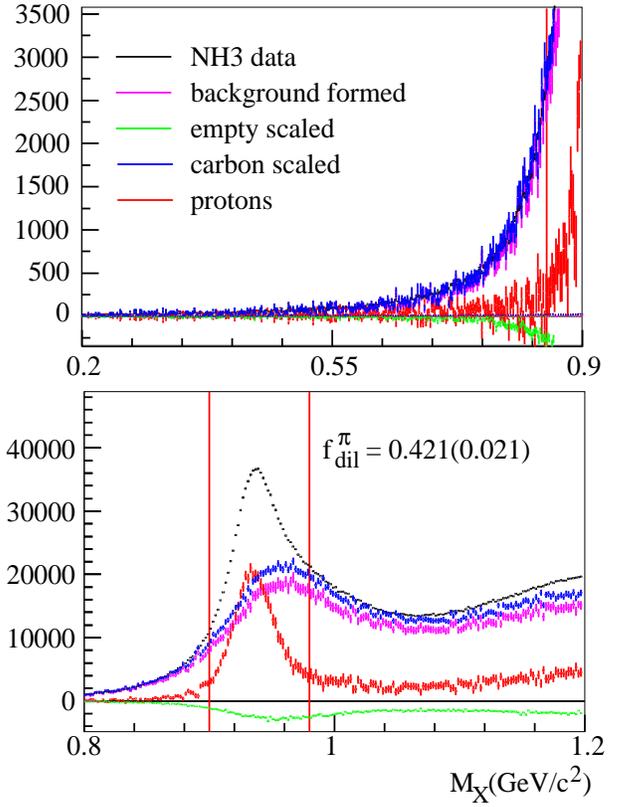}
 \end{center}
 \caption{Missing mass $M_X$ spectrum for deriving the dilution factor for the 
   $ep\to e^\prime \pi^+(n)$ channel. (Top) Missing mass below the neutron mass peak;
   (bottom) missing mass around the neutron mass peak. The data shown are for the 3.0 GeV run period 
using the NH$_3$ long top target.  
Here, the $M_X$ spectrum for the nuclear material (magenta) in the polarized NH$_3$ target 
was constructed using the spectra for the carbon target (blue), the empty target (green), 
with an input packing factor $x=0.65$~cm. 
The nuclear contribution was then subtracted from the NH$_3$ target spectrum (black) 
to give the polarized-proton spectrum (red). The dilution factor was evaluated using the
region around the neutron peak and is shown in the bottom panel with the uncertainty in the bracket. 
The histogram and the dilution uncertainties include both statistical uncertainties and the 
uncertainty in the scaling or packing factors.  
Note that the empty target (green) spectrum is negative, indicating 
we have scaled up the carbon data and then subtracted the extra helium (empty target) 
to reproduce the unpolarized background in NH$_3$. 
Results for the dilution factor is shown in the bottom plot.  
The $M_X$ cuts $(0.90,0.98)$~GeV/$c^2$ used in the dilution and the asymmetry analysis are 
shown by the two red vertical lines. 
}\label{fig:dil_pip_overall}
\end{figure}

In contrast to the dilution for inclusive $P_bP_t$ analysis that has only 
$Q^2$ dependence (Sec.~\ref{sec:ana_pbpt}), the dilution for
exclusive pion production could vary with all four kinematic variables $W$, $Q^2$, 
$\cos\theta^*$ and $\phi^*$~\cite{thesis:DeVita}. To evaluate 
the dilution factor for all four-dimensional bins of ($W$, $Q^2$, $\cos\theta^*$, $\phi^*$),
the yield from the unpolarized material inside the polarized NH$_3$ target was constructed 
using the missing mass spectra from the carbon and the empty targets. Scaling factors
for the carbon and empty target data were calculated 
following a prescription similar to Eqs.~(\ref{eq:eldil_incl})--(\ref{eq:eldil_rEmpt}), but
with the bound-nucleon fraction $B$ replaced by the ratio $Z/A$ (Table~\ref{tab:material}) 
for the $ep\to e'\pi^+(n)$ [($1-Z/A$) for the $en\to e'\pi^-(p)$] channel. 
For NH$_3$ one should use $\frac{Z_{\mathrm{NH}_3}}{A_{\mathrm{NH}_3}}=7/18$ to account for only 
unpolarized protons. We obtain: 
\begin{eqnarray}
 \frac{N_\mathrm{N~in~{NH}_3}}{Q_{\mathrm{NH}3}} &=& 
a \left(\frac{N_{^{12}\mathrm{C}}}{Q_{^{12}\mathrm{C}}}\right) + b \left(\frac{N_\mathrm{empt}}{Q_\mathrm{empt}}\right)~, \label{eq:excldil_ab}
\end{eqnarray}
where
\begin{eqnarray}
  a&=&\frac{\left(\frac{Z_{\mathrm{NH}_3}}{A_{\mathrm{NH}_3}}\rho_{\mathrm{NH}_3}x_{\mathrm{NH}_3}\right)
            + \left(\frac{Z_{w}}{A_{w}}\rho_{w}x_{w}\right) \frac{x_{\mathrm{NH}_3}}{l}
           }
           {\left(\frac{Z_{^{12}\mathrm{C}}}{A_{^{12}\mathrm{C}}}\rho_{^{12}\mathrm{C}}x_{^{12}\mathrm{C}}\right)
            + \left(\frac{Z_{w}}{A_{w}}\rho_{w}x_{w}\right) \frac{x_{^{12}\mathrm{C}}}{l}
           }~,  \label{eq:fillfacta}\\
  b&=& \left(1-\frac{x_{\mathrm{NH}_3}}{l}\right)-\left(1-\frac{x_{^{12}\mathrm{C}}}{l}\right)a~.
               \label{eq:fillfactb}
\end{eqnarray}
Similar to elastic analysis, the value of $b$ from  
Eq.~(\ref{eq:fillfactb}) could be either positive or negative depending on the 
input packing factor.
Figure~\ref{fig:dil_pip_overall} shows the dilution factor evaluation for the 
3.0~GeV data using the NH$_3$ long top target. 

From Eqs.~(\ref{eq:at}) and (\ref{eq:aet}) one can see that the 
uncertainties in $P_bP_t$ and $f_\mathrm{dil}^{\pi}$ should be evaluated at the same time 
because both depend on the packing factor. 
Table~\ref{tab:PbPtfdil} shows all $P_bP_t$ and dilution results for the packing factor
range used in the elastic $P_bP_t$ analysis. For each setting of beam energy and 
target, we varied the packing factor by one standard deviation and evaluated 
$P_bP_t$ and $f_\mathrm{dil}^{\pi^+}$. We used
the observed difference in the product $P_bP_tf_\mathrm{dil}^{\pi^+}$ as the uncertainty from  
the packing factor, labeled as $P_bP_tf_\mathrm{dil}^{\pi^+}\pm (p.f.)$. For the total uncertainty
$\frac{\Delta(P_bP_tf_\mathrm{dil})}{P_bP_tf_\mathrm{dil}}$ (total), we added the following terms 
in quadrature:
(1) statistical uncertainty of inclusive elastic events used in the $P_bP_t$ analysis; 
(2) statistical uncertainty of the carbon and empty target counts used to calculate
the dilution factor for inclusive elastic events; 
(3) statistical 
uncertainty in the exclusive $ep\to e'\pi^+(n)$ channel from limited statistics of carbon and 
empty target data $f_\mathrm{dil}^{\pi^+}\pm$(stat.); and (4) the observed variation in 
$P_bP_tf_\mathrm{dil}^{\pi^+}$ when the input packing factor was varied within its uncertainty. 
The resulting total uncertainties on $P_bP_tf_\mathrm{dil}^{\pi^+}$ were used for the evalulation 
of the uncertainty of the double-spin asymmetry $A_{LL}$. 
For the target asymmetry $A_{UL}$, the uncertainty was evaluated by combining the 
uncertainty of $P_bP_tf_\mathrm{dil}^{\pi^+}$ and the uncertainty of the M\o ller measurements 
on the beam polarization. The uncertainty from the polarizations and the dilution is the largest 
systematic uncertainty of the present analysis.

\begin{table*}[!t]
 \caption{Dilution factor $f_\mathrm{dil}^{\pi^+}$  
and the product $P_bP_tf_\mathrm{dil}^{\pi^+}$ for the exclusive $\pi^+$ channel. 
The $P_bP_t$ results extracted from 
inclusive elastic scattering, described in Sec.~\ref{sec:ana_pbpt}, and their 
uncertainties are also shown. For $P_bP_t$, the three errors are from 
statistical uncertainty of the elastic events, the statistical uncertainty of the  
carbon and empty target counts used to calculate
the dilution factor for inclusive elastic analysis, 
and the uncertainty of the packing factor. 
$P_bP_t$ values from M\o ller and NMR measurements are shown for comparison, 
although the NMR measurements are unreliable as decribed in Sec.~\ref{sec:target}. 
The products $P_bP_tf_\mathrm{dil}$ are used to correct
the exclusive channel asymmetries. 
The total uncertainties in $P_bP_tf_\mathrm{dil}$ include uncertainties of
$P_bP_t$, statistical uncertainties of $f_\mathrm{dil}^{\pi^+}$, and the uncertainties 
from the packing factor (p.f.), all added 
in quadrature. These total uncertainties will be used as systematic uncertainties on the 
extracted exclusive channel asymmetries.}\label{tab:PbPtfdil}
 \begin{tabular}{|c|c|c|c|c|c|c|c|}\hline\hline
   $E_\mathrm{beam}$ & Target    & p.f. & $(P_bP_t)_{el}$ & M\o ller& $f_\mathrm{dil}^{\pi^+}\pm$(stat.)$\pm$(p.f.) 
  & $P_bP_tf_\mathrm{dil}$ &$\frac{\Delta(P_bP_tf_\mathrm{dil})}{P_bP_tf_\mathrm{dil}}$ \\
    (GeV)          &(NH$_3$)  & (cm) &    & $\times$ NMR & &       & (total)              \\\hline
{3.0}& top & $0.65\pm 0.05$ & $0.614\pm 0.006\pm 0.015\pm 0.045$ & $0.620$ & $0.424\pm 0.021\pm 0.013$ & $0.260$ & $7.0\%$ \\
\hline
\multirow{2}{*}
{2.3}& top & $0.65\pm 0.05$ & $0.597\pm 0.006\pm 0.021\pm 0.028$ & $0.551$ & $0.476\pm 0.021\pm 0.011$ & $0.284$ & $6.2\%$  \\
\cline{2-8}
     &short& $0.30\pm 0.05$  & $0.560\pm 0.009\pm 0.026\pm 0.067$ & $0.601$ & $0.322\pm 0.017\pm 0.021$ & $0.180$ & $9.0\%$  \\
\hline
\multirow{2}{*}
{2.0}& top & $0.65\pm 0.05$& $0.605\pm 0.004\pm 0.016\pm 0.030$ & $0.545$ & $0.495\pm 0.020\pm 0.010$ & $0.299$ & $5.7\%$  \\
\cline{2-8}
    &bottom& $0.65\pm 0.05$ & $0.636\pm 0.019\pm 0.016\pm 0.031$ & $0.560$ & $0.484\pm 0.021\pm 0.010$ & $0.308$ & $6.4\%$  \\
\hline
\multirow{3}{*}
{1.3}& top & $0.70\pm 0.05$  & $0.571\pm 0.003\pm 0.009\pm 0.033$ & $0.509$ & $0.494\pm 0.019\pm 0.010$ & $0.282$ & $5.7\%$  \\
\cline{2-8}
    &bottom& $0.70\pm 0.05$& $0.535\pm 0.003\pm 0.010\pm 0.028$ & $0.458$ & $0.493\pm 0.019\pm 0.010$ & $0.264$ & $5.5\%$ \\
\cline{2-8}
    &short & $0.30\pm 0.05$ & $0.552\pm 0.010\pm 0.030\pm 0.060$ & $0.581$ & $0.383\pm 0.016\pm 0.014$ & $0.211$ & $10.2\%$ \\
\hline
{1.1}&bottom& $0.75\pm 0.10$ & $0.568\pm 0.002\pm 0.007\pm 0.080$ & $0.563$ & $0.496\pm 0.020\pm 0.020$ & $0.282$ &  $11.1\%$\\
%
%
\hline
 \end{tabular}
\end{table*}

The uncertainty in the input packing factor of Table~\ref{tab:PbPtfdil} was 
checked using not only the $W$ spectrum of elastic events (as described in 
Sec.~\ref{sec:ana_pbpt}), but also the dilution factor of the 
$en\to e^\prime \pi^-(p)$ channel analyzed using a similar prescription as 
Eqs.~(\ref{eq:excldil_ab})--(\ref{eq:fillfactb}). The 
dilution factor of the $\pi^-(p)$ channel should be consistent with zero 
in all kinematic bins. 
Overall, the lower bound in the packing factor was cross-checked between the 
$en\to e^\prime \pi^-(p)$ dilution result and the elastic $W$ spectrum, and
the upper bound in the packing factor was determined always by the 
elastic $W$ spectrum. 

\begin{figure}[!htp]
  \includegraphics[width=0.35\textwidth]{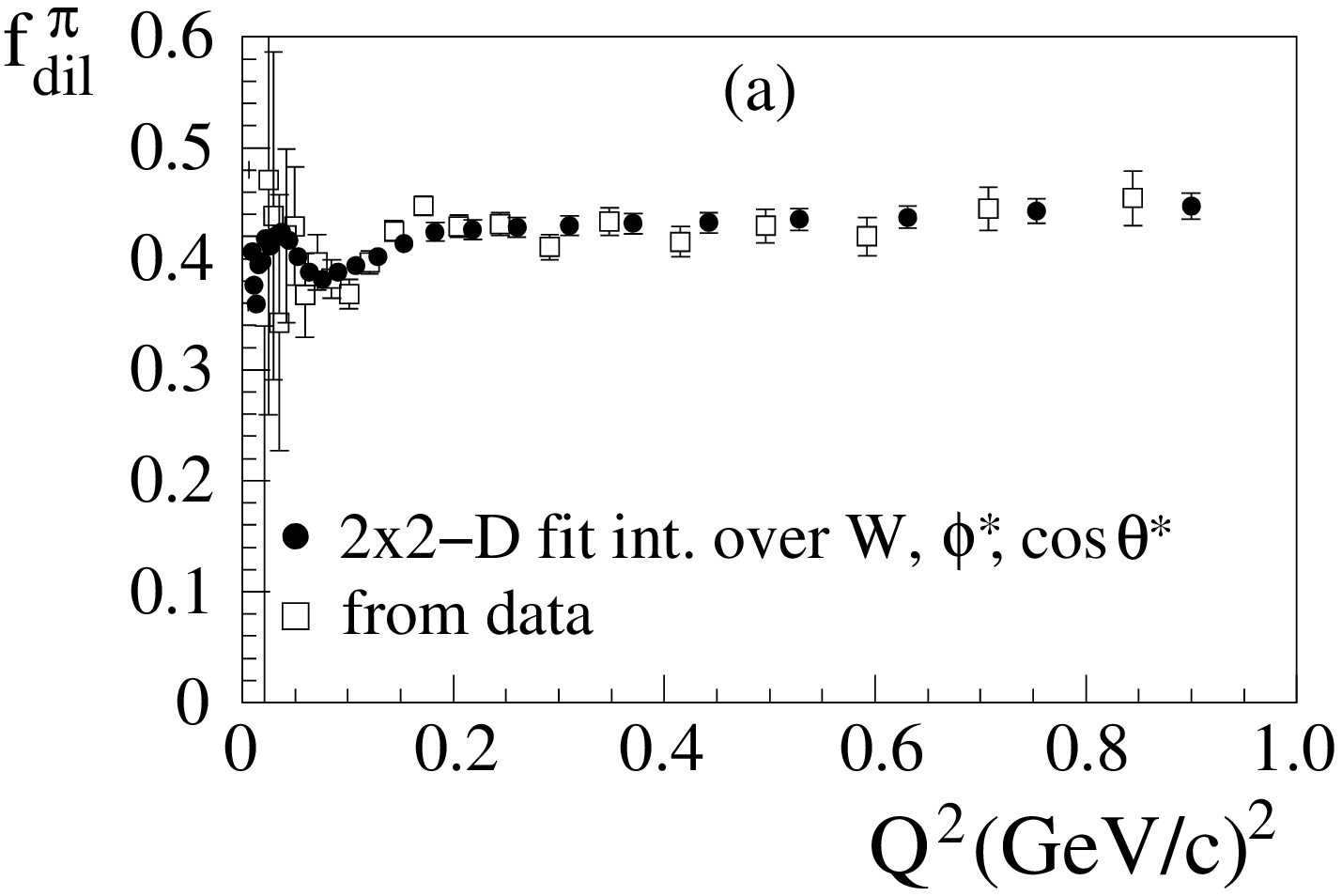}\\
  \includegraphics[width=0.35\textwidth]{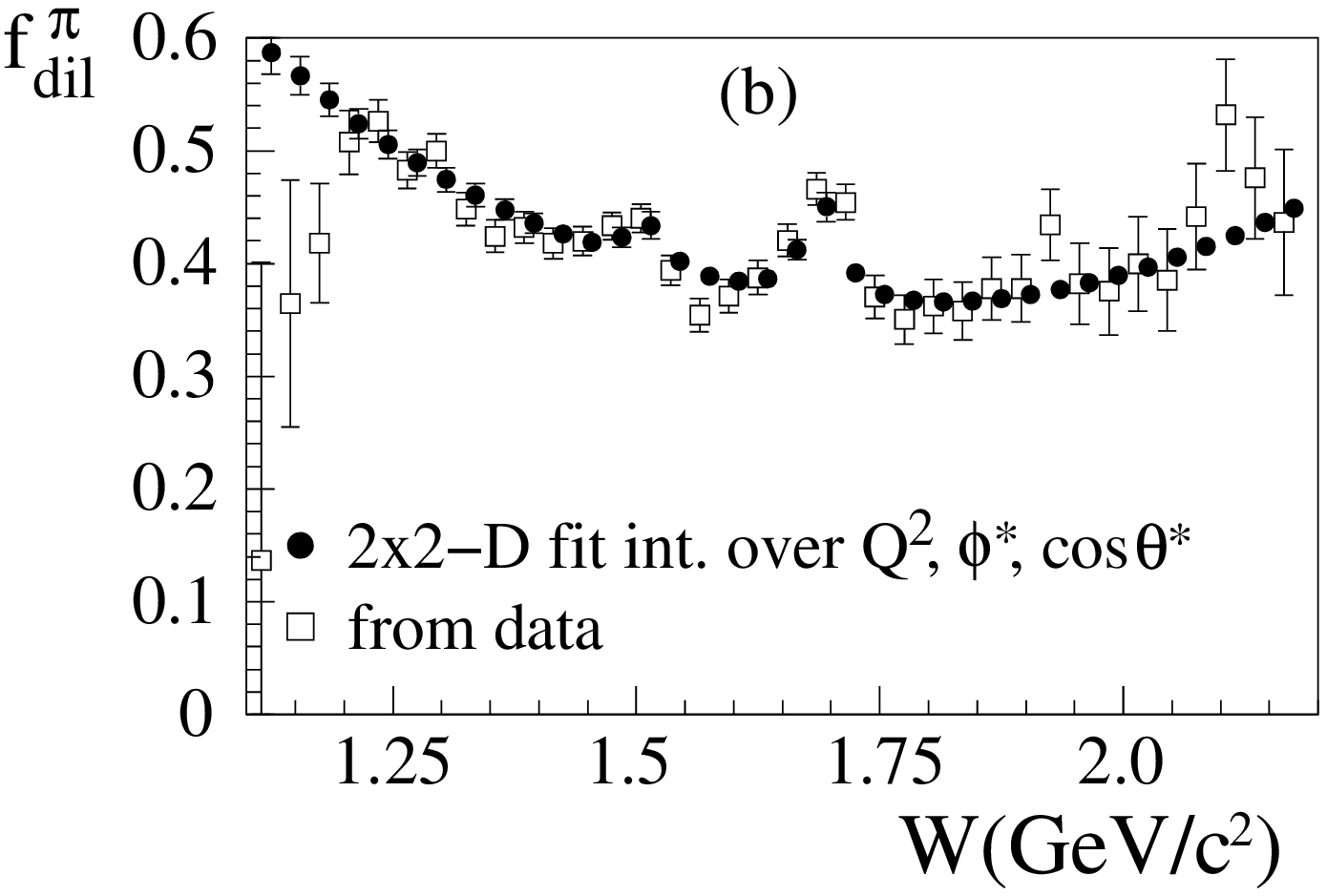}\\
  \includegraphics[width=0.35\textwidth]{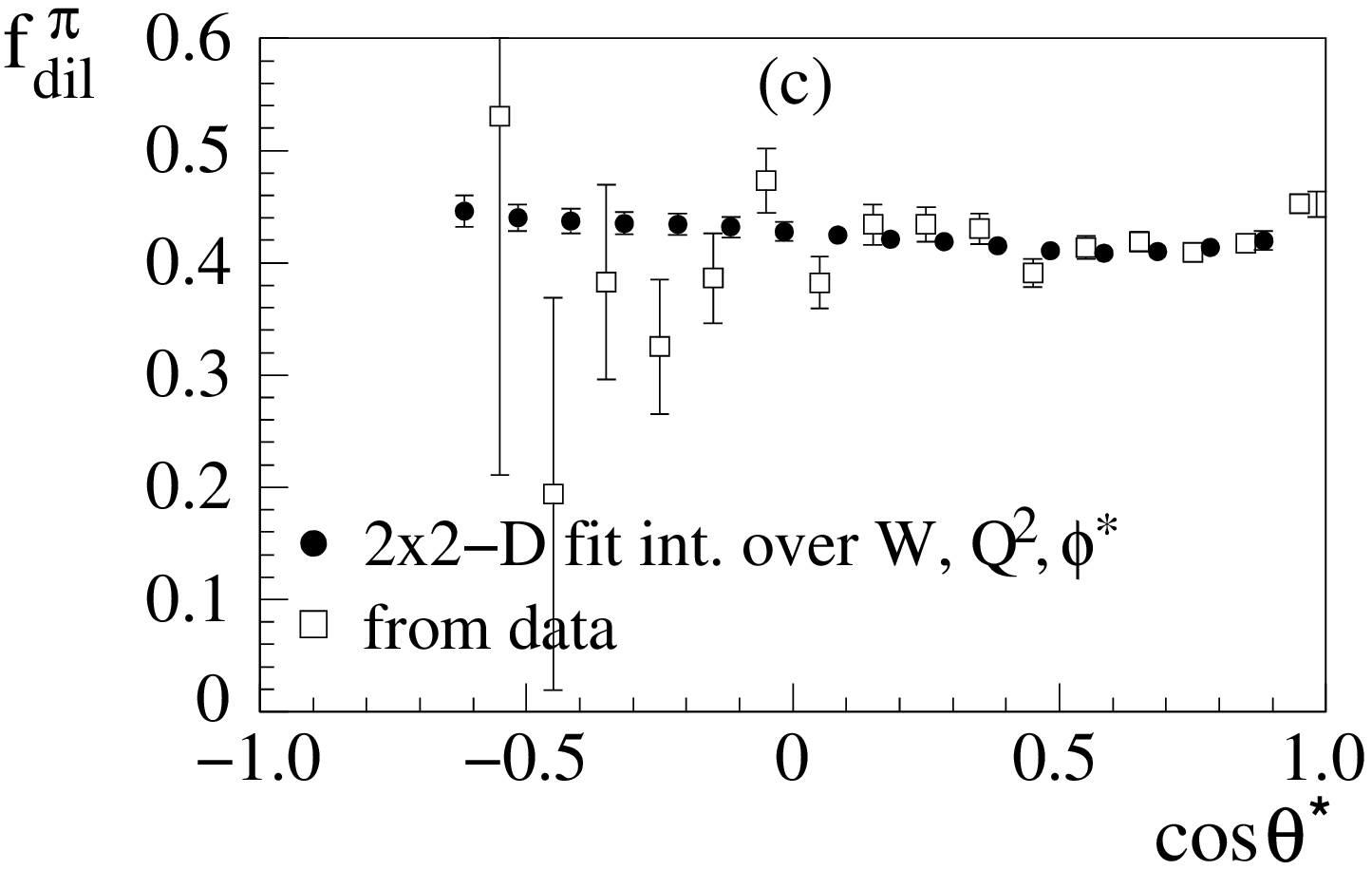}\\
  \includegraphics[width=0.35\textwidth]{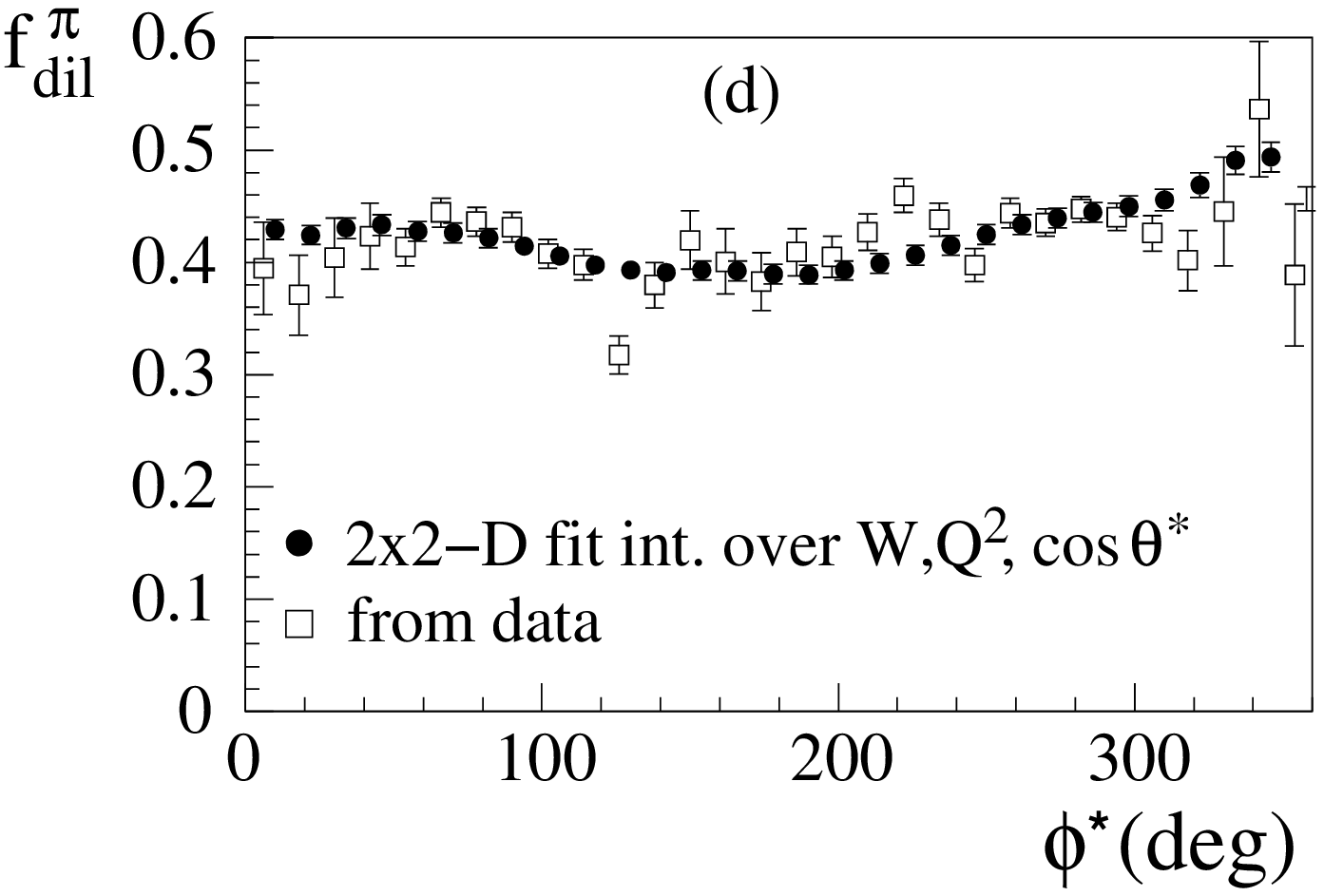}
  \caption{Dependence of dilution on: (a) $Q^2$, (b) $W$, (c) $\cos\theta^*$, and (d) $\phi^*$,
    for the 3.0~GeV NH$_3$ long top target, $ep\to e^\prime \pi^+(n)$ channel, obtained directly
    from the data (open squares) and from multiplying the two 2D fits of
    Eqs.~(\ref{eq:2dfit_q2w}) and (\ref{eq:2dfit_thetaphi}) then integrating over three of the four 
    variables (solid circles). The error bars for the dilution extracted from data are
    statistical only.}\label{fig:dil_pip_2dfits}
\end{figure}

The kinematics dependence of the dilution factor on $Q^2$, $W$, and the pion center-of-mass 
angles $\theta^*$ and $\phi^*$ have been studied, and multi-dimensional fits of the
dependence were performed. The limited statistics 
of the carbon and the empty target data prevented fitting the 
$(Q^2,W,\cos\theta^*,\phi^*)$ dependence simultaneously. Instead, two 
bi-dimensional fits were used, one for the $(Q^2,W)$ dependence and one for the 
$(\cos\theta^*,\phi^*)$ dependence, with the following {\it ad hoc} parametrizations: 
 \begin{eqnarray}
        f_1 &=& p_0\left[1+p_1(Q^2)+p_2(Q^2)^2\right] \nonumber\\
          && \times \left[1+p_3(W-1.8)+p_4(W-1.8)^2\right]\nonumber\\
          && \times \left[1+\frac{p_5}{(W^2-1.50^2)^2+1.50^2\times 0.05^2}\right]\nonumber\\
          && \times \left[1+\frac{p_6}{(W^2-1.68^2)^2+1.68^2\times 0.05^2}\right] \label{eq:2dfit_q2w}
      \end{eqnarray}
where $W$ is in GeV$/c^2$ and 
      \begin{eqnarray}
        f_2  &=& p_0^\prime \times \left[1+\frac{p_7}{1-\cos\theta^*}\right] \nonumber\\
          && \times \left[1+p_8 \sin\phi^*+p_9\cos\phi^*\right]~.\label{eq:2dfit_thetaphi}
      \end{eqnarray}
The resulting two fits were then multiplied to give the overall $2\times 2$-dimensional 
fit for $f_\mathrm{dil}^{\pi}(W,Q^2,\cos\theta^*,\phi^*)$.
To check the validity of the fit, the results from 
$f_\mathrm{dil}^{\pi}(W,Q^2,\cos\theta^*,\phi^*)$ were integrated over three of the four
variables, and then compared with the dilution extracted directly from data
binned in the fourth variable. This comparison is shown in 
Fig.~\ref{fig:dil_pip_2dfits}.  One can see that the dilution 
factors obtained from this method agree with data very well. The $2\times 2$-dimensional
fit $f_\mathrm{dil}^{\pi}(W,Q^2,\cos\theta^*,\phi^*)$ was used to correct the 
asymmetries $A_{UL}$ and $A_{LL}$ for the specific $W,Q^2,\cos\theta^*,\phi^*$ bin using 
Eqs.~(\ref{eq:at}) and (\ref{eq:aet}). 

\subsection{Effect of nitrogen polarization on the asymmetry}\label{sec:ana_15N}

The $^{15}$N in the NH$_3$ target is polarizable and can affect the measured asymmetry. 
In this section we estimate this effect and show that it is negligible. Therefore 
no correction was made to the extracted exclusive channel asymmetries.

The nitrogen polarization in $^{15}$NH$_3$ can be estimated based on the equal spin 
temperature (EST) prediction~\cite{Crabb:1995xi}:
\begin{eqnarray}
  P(^{15}\mathrm{N}) = \tanh\frac{\mu_{^{15}\mathrm{N}}B}{kT_S},
  P(\mathrm{H}) = \tanh\frac{\mu_{p}B}{kT_S},
\end{eqnarray}
where $\mu_{^{15}\mathrm{N}}$ and $\mu_p$ are the magnetic moments of the $^{15}$N and the 
proton, respectively, $B$ is the magnetic field of the target, $k$ is the Boltzmann 
constant and $T_S$ is the spin 
temperature that describes the Boltzmann distribution of spins inside the target. 
The EST prediction was demonstrated to apply to the $^{15}$N and H of the ammonia molecule
by several experiments starting with the Spin Muon Collaboration~\cite{Adeva:1997kz}.
The SLAC E143 collaboration performed an empirical fit and showed~\cite{Abe:1998wq}:
\begin{eqnarray}
  P_{^{15}\mathrm{N}} = 0.136\vert P_p\vert - 0.183\vert P_p\vert^2 + 0.335\vert P_p\vert^3~,
\end{eqnarray}
which gives $P_{^{15}\mathrm{N}} \approx -15\%$ when $P_p=90\%$ and $P_{^{15}\mathrm{N}} \approx -8.8\%$ when $P_p=70\%$.
The $^{15}$N polarization is carried by the unpaired proton and its effect relative
to the three free protons in NH$_3$ is
\begin{eqnarray}
  \Delta P&=&\frac{1}{3}\left(-\frac{1}{3}\right)P(^{15}\mathrm{N})~,\label{eq:15NvsH}
\end{eqnarray}
where the additional factor of $-1/3$ comes from the wavefunction of the unpaired
proton in the $^{15}$N~\cite{RondonAramayo:1999da}. 
The effect on the asymmetry from the polarized proton in the $^{15}$N is thus 
at the $(1-2)\%$ level, 
and is negligible compared to the statistical uncertainty
of the asymmetry and the systematic uncertainty from the polarizations and the dilution 
factor.

\subsection{Acceptance corrections}~\label{sec:accp_corr}
When studying how the asymmetries vary with very small bins in all 
four kinematic variables -- the electron's $Q^2$, $W$ and the pion's center-of-mass
angles $\theta^*$ and $\phi^*$ -- the effect of the detector acceptance and efficiency 
in principle cancel and therefore do not affect the interpretation of the asymmetry results. 
The effect of acceptance only becomes relevant when integration of the asymmetry over 
a subset of these four variables is necessary, which is the case for all results presented 
in Sec.~\ref{sec:results}. 

For results presented in Sec.~\ref{sec:results}, we evaluated the acceptance of each 
bin based on acceptance cuts for both electrons and pions. The acceptance correction was then applied
on an event-by-event basis: instead of using the measured counts $N_{\ua\Ua,\ua\Da,\da\Ua,\da\Da}$, 
where each event counts as $1$, we first divided $1$ by the acceptance of that particular 
event, then the sum was taken and used as $N_{\ua\Ua,\ua\Da,\da\Ua,\da\Da}$ in the formula from 
Sec.~\ref{sec:ana_procedure}, Eqs.~(\ref{eq:ae})-(\ref{eq:aet}).
The asymmetries extracted this way were integrated over certain kinematic ranges and
compared directly with theoretical predictions. 
Zero-acceptance bins could not be corrected this way when integrating the data. When 
integrating the theoretical calculations, we excluded bins where there were no data, 
and thus removed the zero-acceptance bins from the theory curves as well. 

\subsection{Radiative corrections}
Radiative corrections were calculated for both $A_{UL}$ and $A_{LL}$ using the code EXCLURAD~\cite{EXCLURAD}
and the MAID2007 model~\cite{Drechsel:2007if}. It was found that overall the correction is fairly small and typically no larger
than 0.03. Considering the size of the statistical uncertainty of the 
measurement, radiative corrections were not applied to the asymmetries, but
rather are quoted as a systematic uncertainty of $\Delta A=\pm 0.03$ throughout the 
accessed kinematics.

\subsection{Summary of all systematic uncertainties}\label{sec:all_syst}
The systematic uncertainty of the $\vec e\vec p\to e'\pi^+(n)$ exclusive channel is dominated by 
that from the product $P_bP_tf_\mathrm{dil}^{\pi^+}$, shown in Table~\ref{tab:PbPtfdil}. 
The uncertainty of $P_bP_tf_\mathrm{dil}^{\pi^+}$ takes into account the
uncertainties in the target packing factor, as well as the thickness and density of 
various materials in the target. Other non-neglible systematic uncertainties include a 
relative $\pm (1-2)\%$ due to the $^{15}$N in NH$_3$ and a $\pm 0.03$ due to radiative 
corrections.
Adding these uncertainties in quadrature, we arrive at 
Table~\ref{tab:syst_summary} for our asymmetry results. 
For the asymmetry $A_{UL}$, one does not need to normalize by $P_b$. We relied on the
elastic $P_bP_t$ results and combined in quadrature their uncertainties with the 
uncertainty in the M\o ller polarization to obtain the 
uncertainty on $P_t$ alone. 
\begin{table}
 \caption{Summary of systematic uncertainties from the target and beam polarizations and the 
dilution factor for different beam and target combinations.  The $(1-2)\%$ relative uncertainty
from $^{15}$N and the $\pm 0.03$ absolute uncertainty from radiative corrections must
be added in quadrature to the values here to obtain the total systematic uncertainty.
}\label{tab:syst_summary}
\begin{center}
 \begin{tabular}{c|c|c|c}\hline\hline
   $E_\mathrm{beam}$ & Target   & $\Delta A_{UL}/A_{UL}$ & $\Delta A_{LL}/A_{LL}$ \\
    (GeV)          &(NH$_3$)   & (syst) & (syst) \\\hline
{3.0}& top    & $7.0\%$ & $7.0\%$ \\
\hline
\multirow{2}{*}{2.3}& top   & $6.2\%$ & $6.3\%$ \\
                    & short & $9.0\%$ & $9.0\%$ \\\hline
{2.0}& top   & $5.7\%$ & $5.8\%$  \\\hline
\multirow{2}{*}{1.3}& top   & $5.7\%$ & $5.9\%$ \\
             & bottom & $5.5\%$ & $5.7\%$ \\\hline
{1.1}&       bottom &  $11.1\%$ & $11.2\%$\\\hline\hline
\end{tabular}
\end{center}
\end{table}

\section{Asymmetry Results}\label{sec:results}
Results for the target asymmetry $A_{UL}$ and the double-spin asymmetry $A_{LL}$ 
are available on a four-dimensional grid of $Q^2$, $W$, $\cos\theta^*$, and 
$\phi^*$. There are $42$ $Q^2$ bins logarithmically spaced between $0.00453$
and $6.45$~(GeV/$c$)$^2$, $38$ $W$ bins between $1.1$ and $2.21$~GeV$/c^2$, 
$30$ $\phi^*$ bins between $0$ and $360^\circ$, 
and 20 $\cos\theta^*$ bins between $-1$ and $1$. This binning 
scheme is referred to as ``asymmetry bins''. 
To allow a meaningful comparison with theoretical calculations, we 
integrated the data over 3 $Q^2$ bins, 8 $W$ bins, 5 $\phi^*$ bins and
5 $\cos\theta^*$ bins. These will be referred to as ``combined bins'' 
hereafter. The resulting combined $W$ bins are $(1.1, 1.34)$, $(1.34,1.58)$ 
and $(1.58,1.82)$~GeV$/c^2$, allowing an examination of the first, the second, 
and the third nucleon resonance regions, respectively. 

The method of integrating the data for the combined bins
was built upon the acceptance correction described in Sec.~\ref{sec:accp_corr}: 
To correct for the acceptance, each event in the asymmetry bin was
divided by the acceptance of that particular event, then summed to be used as
$N_{\ua\Ua,\ua\Da,\da\Ua,\da\Da}$ in Eqs.~(\ref{eq:ae})--(\ref{eq:aet}).
To integrate from asymmetry bins into combined bins, these acceptance-corrected
$N_{\ua\Ua,\ua\Da,\da\Ua,\da\Da}$ from each asymmetry bin was summed, and used
as the combined $N_{\ua\Ua,\ua\Da,\da\Ua,\da\Da}$ to evaluate the asymmetries for the combined bin. 
Using this method, the integrated asymmetries are direct reflections of the
ratio of the physical cross sections integrated over the combined bin except 
for regions that had zero acceptance. 
To compare with theory, we calculated the cross sections 
$\sigma_{t,et,0}$ for each asymmetry bin, 
then summed the calculated cross sections over combined bins except for asymmetry bins
where there was no data (zero acceptance). The ratio of the summed cross 
sections [Eqs.~(\ref{eq:AULdef}) and (\ref{eq:ALLdef})] was taken 
as the calculated asymmetry for the combined bin. 
In the following we will present some representative results.

\subsection{Results on target asymmetry $A_{UL}$}
Figure~\ref{fig:at_w} shows, in increasing $Q^2$ ranges, the $A_{UL}$ results as
a function of $W$ for three $\phi^*$ bins $(120^\circ,180^\circ)$, $(180^\circ,240^\circ)$,
$(240^\circ,300^\circ)$, and integrated over $0.5<\cos\theta^*<1.0$.
Results for the $\phi^*=(0^\circ,60^\circ)$ and $(300^\circ,360^\circ)$ have less statistics
and are not shown. Results for the $\phi^*=(60^\circ,120^\circ)$ bin have comparable
statistics as Fig.~\ref{fig:at_w} but are not shown here for brevity. 
In general, we see that the agreement between these $A_{UL}$ results and the four calculations,
MAID2007 (solid)~\cite{Drechsel:2007if},
JANR (dashed)~\cite{Aznauryan:2009mx},
SAID (dash-dotted)~\cite{Arndt:2009nv}, 
and DMT2001 (dotted)~\cite{Kamalov:1999hs},
is very good in the $W<1.5$~(GeV/$c^2$) region, but for the region
$1.5<W<1.8$~(GeV/$c^2$), all four calculations differ from each other and
none agrees well with data, although the MAID2007 curve (solid)
approximates the data better than the other three. 

To study these results further for different $W$ regions,  
we show in Fig.~\ref{fig:at_phi} $A_{UL}$ results as
a function of $\phi^*$ for three $W$ ranges and between $Q^2=0.0187$ and $0.452$~(GeV/$c$)$^2$. 
Results for lower $Q^2$ ranges, down to $0.00646$~(GeV/$c$)$^2$, are available from the 1.1~GeV
data but only cover $1.2<W<1.5$~(GeV/$c^2$) and thus are not presented here.
From Fig.~\ref{fig:at_phi}, for the lower two $W$ bins $(1.12, 1.34)$ and
$(1.34, 1.58)$~GeV$/c^2$, the four calculations provide similar predictions and all agree with
data. But for the $W=(1.58, 1.82)$~GeV$/c^2$ region, only the MAID2007 (solid) and the DMT2001 (dotted)
calculations provide the correct sign, and MAID2007 approximates the data better than the other three 
although it does not agree with data perfectly. 
It is clear that all four calculations can be improved in 
the $W>1.58$~GeV$/c^2$ region throughout the $Q^2$ range shown.

\subsection{Results on the double-spin asymmetry $A_{LL}$}

Figure~\ref{fig:aet_w} shows the double-spin asymmetry $A_{LL}$ results as  
a function of $W$ for eight $Q^2$ bins, three $\phi^*$ bins, and integrated over 
$\cos\theta^*=(0.5,1.0)$.
These results are compared with four calculations: MAID2007 (solid)~\cite{Drechsel:2007if},
JANR (dashed)~\cite{Aznauryan:2009mx},
SAID (dash-dotted)~\cite{Arndt:2009nv}, 
and DMT2001 (dotted)~\cite{Kamalov:1999hs}. 
Note that our definiton for $A_{LL}$ has the opposite sign 
from theories; see Sec.~\ref{sec:intro_formula}.
Results for the $\phi^*=(0^\circ,60^\circ)$ and $(300^\circ,360^\circ)$ bins have less statistics
and are not shown. Results for the $\phi^*=(60^\circ,120^\circ)$ bin have comparable
statistics as Fig.~\ref{fig:at_w} but are not shown here for brevity. 
Overall the data agree very well with all four calculations. 
For all $\phi^*$ bins, the sign of $A_{LL}$ in the region of the $N(1520)3/2^-$ and the
$N(1680)5/2^+$ is positive in the high $Q^2$, but start to cross or approach zero in the lower
$Q^2$ bin, within $(0.0919,0.156)$~(GeV/$c$)$^2$ 
for $N(1520)3/2^-$ and within $Q^2=(0.266,0.452)$~(GeV/$c$)$^2$ for $N(1680)5/2^+$, respectively. 
This is in agreement with the 
suggestion in Sec.~\ref{sec:motivation} that $A_{LL}$ turns to positive at 
high $Q^2$ values from helicity conservation, but may become negative near the real photon point. 


\section{Summary}
We present here data on the target and double-spin asymmetry $A_{UL}$ and $A_{LL}$
on the $\vec e\vec p\to e\pi^+(n)$ channel using data taken on a polarized NH$_3$ target, 
from the EG4 experiment using CLAS in Hall B of Jefferson Lab. These data
have reached a low $Q^2$ region from $0.0065$ to $0.35$ (GeV/$c$)$^2$ that was not 
accessed previously. They suggest a transition in $A_{LL}$ from positive at higher 
$Q^2$ to negative values below $Q^2\approx 0.1$~(GeV/$c$)$^2$ in the region $1.5<W<1.7$~GeV$/c^2$, 
in agreement with both previous data from CLAS
(high $Q^2$)~\cite{EG1b-Pierce-thesis,Bosted:2016leu} and the real photon data at $Q^2=0$.
Our results show that while all model calculations agree well with $A_{LL}$, 
in general there is room for improvements for $A_{UL}$ in the high-mass 
resonance region $W>1.58$~GeV$/c^2$ where predications from various models
differ significantly.

\begin{acknowledgments}

The authors gratefully acknowledge the work of Jefferson Lab staff 
in the Accelerator and Physics Divisions that 
resulted in the successful completion of the experiment. This work was supported by:
the U.S. Department of Energy (DOE), 
the U.S. National Science Foundation,
the U.S. Jeffress Memorial Trust; 
the United Kingdom's Science and Technology Facilities Council (STFC)
under grant numbers 
ST/L005719/1 
and GR/T08708/01; 
the Italian Istituto Nazionale di Fisica Nucleare;
the French Institut National de Physique Nucl\'{e}aire et de Physique des Particules,
the French Centre National de la Recherche Scientifique;
and the National Research Foundation of Korea. 
This material is based upon work supported by the U.S. Department of Energy, Office of Science, Office of Nuclear Physics under contract DE-AC05-06OR23177.

\end{acknowledgments}

\begin{widetext}

\begin{figure}[!htp]
 \includegraphics[width=0.9\textwidth]{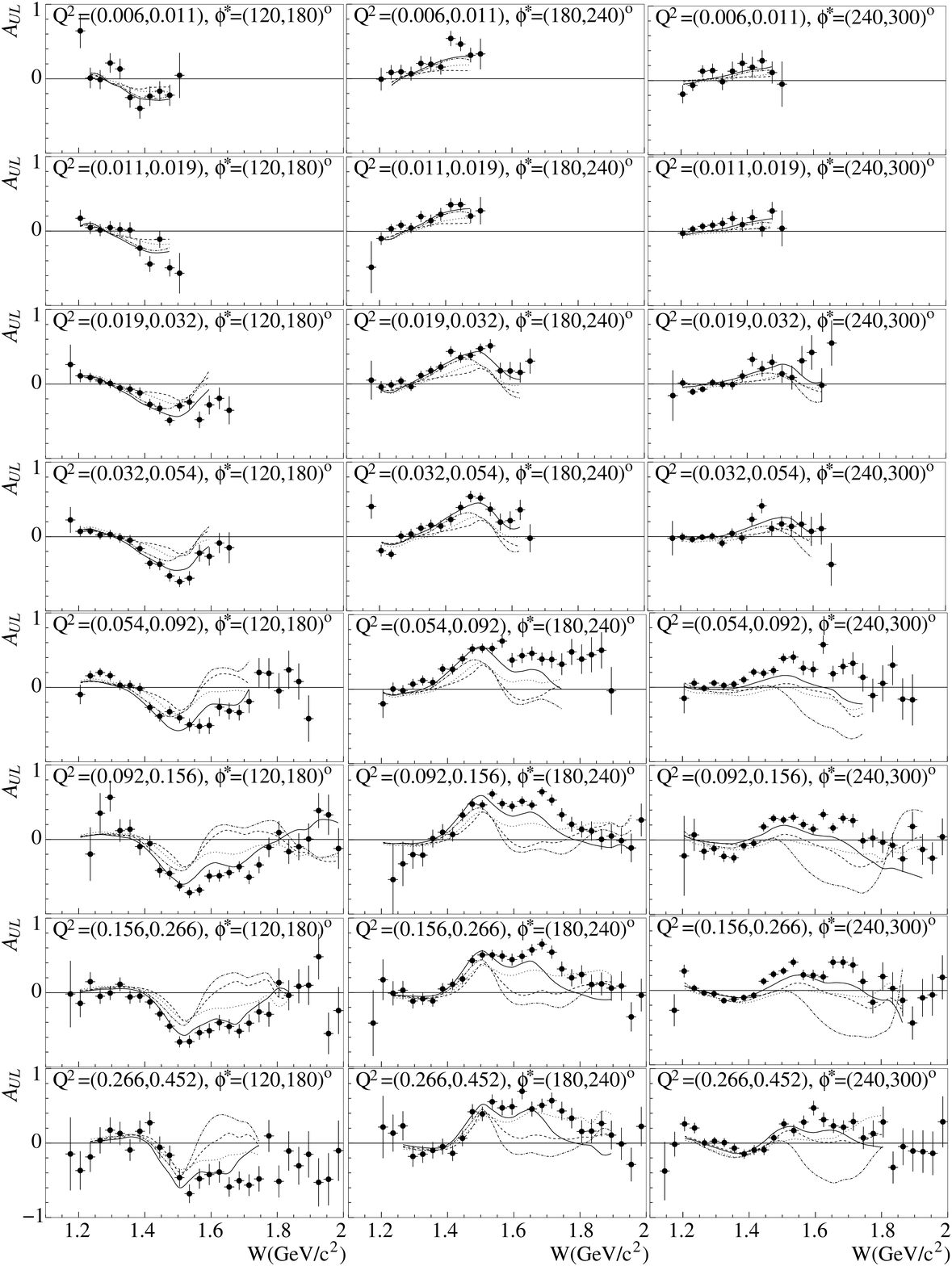}
 \caption{Results on the target spin symmetries $A_{UL}$ for the $\vec e\vec p\to e\pi^+(n)$ channel 
   as a function of the invariant mass $W$ in GeV$/c^2$, integrated over $\cos\theta^*=(0.5,1.0)$,
   in increasing $Q^2$ ranges and three 60$^\circ$ $\phi^*$ bins.
From top to bottom the $Q^2$ bins are
$(0.00646,0.0110)$ and 
$(0.0110,0.0187)$ (1.1 GeV NH$_3$ long bottom target), 
$(0.0187,0.0317)$ and 
$(0.0317,0.054)$ (1.3~GeV NH$_3$ long top target), 
$(0.054,0.0919)$ (2.0~GeV NH$_3$ long top target), 
$(0.0919,0.156)$, 
$(0.156,0.266)$, and
$(0.266,0.452)$~(GeV/$c$)$^2$ (3.0~GeV NH$_3$ long top target).
From left to right the $\phi^*$ bins are
$\phi^*=(120^\circ,180^\circ)$, 
$(180^\circ,240^\circ)$ and 
$(240^\circ,300^\circ)$. 
In each panel, the horizontal scale is from 1.1 to 2~GeV$/c^2$ in $W$ and the vertical scale is 
from $-1$ to $1$.
Data are compared to four calculations: 
MAID2007 (solid)~\cite{Drechsel:2007if}, JANR (dashed)~\cite{Aznauryan:2009mx},
SAID (dash-dotted)~\cite{Arndt:2009nv}, 
and DMT2001 (dotted)~\cite{Kamalov:1999hs}.
}\label{fig:at_w}
\end{figure}

\begin{figure}[!htp]
 \includegraphics[width=0.9\textwidth]{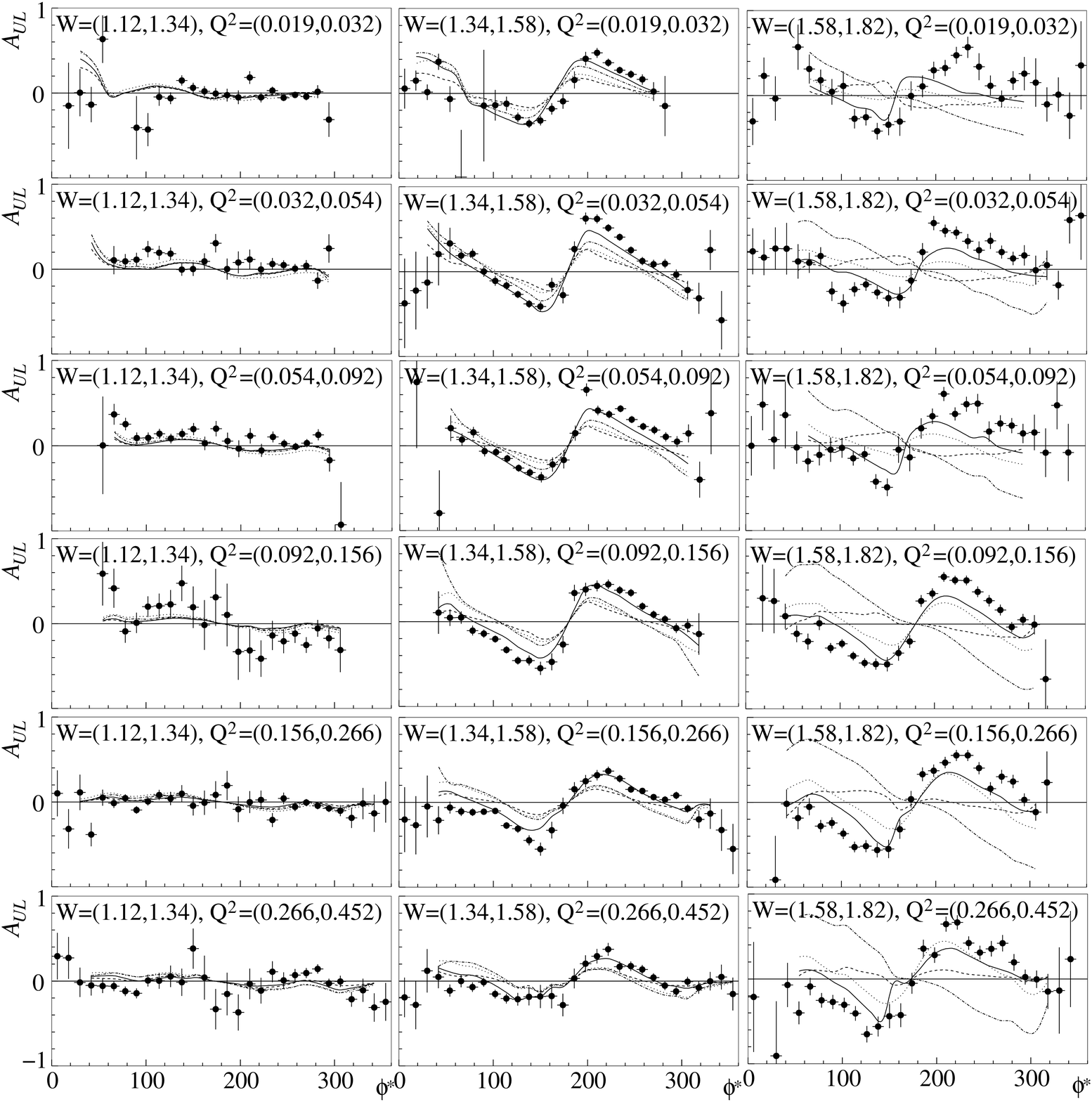}
 \caption{Results on $A_{UL}$ for the $\vec e\vec p\to e\pi^+(n)$ channel 
as a function of azimuthal angle $\phi^*$, integrated over $\cos\theta^*=(0.5,1.0)$, 
for six $Q^2$ bins and three $W$ bins. 
From top to bottom the six $Q^2$ bins are: 
$Q^2=(0.0187,0.0317)$ [1.3 NH$_3$ long target for $W=(1.12,1.34)$ and $(1.34,1.58)$~GeV$/c^2$,
and 2.0 NH$_3$ long top target for $W=(1.58,1.82)$~GeV$/c^2$]; 
$(0.156,0.266)$ and  
$(0.266,0.452)$~(GeV/$c$)$^2$ (2.0 GeV NH$_3$ long top target);
$(0.0919,0.156)$, 
$(0.156,0.266)$ and
$(0.266,0.452)$~(GeV/$c$)$^2$ (3.0 GeV NH$_3$ long top target); 
from left to right the $W$ bins are:
$W=(1.12,1.34)$,
$(1.34,1.58)$,
$(1.58,1.82)$~GeV$/c^2$. 
In each panel, the horizontal scale is from 0 to $360^\circ$ in $\phi^*$ and the vertical scale is 
from $-1$ to $1$.
Data are compared to four calculations: 
MAID2007 (solid)~\cite{Drechsel:2007if}, JANR (dashed)~\cite{Aznauryan:2009mx},
SAID (dash-dotted)~\cite{Arndt:2009nv}, 
and DMT2001 (dotted)~\cite{Kamalov:1999hs}.
}\label{fig:at_phi}
\end{figure}

\begin{figure}[!htp]
 \includegraphics[width=0.9\textwidth]{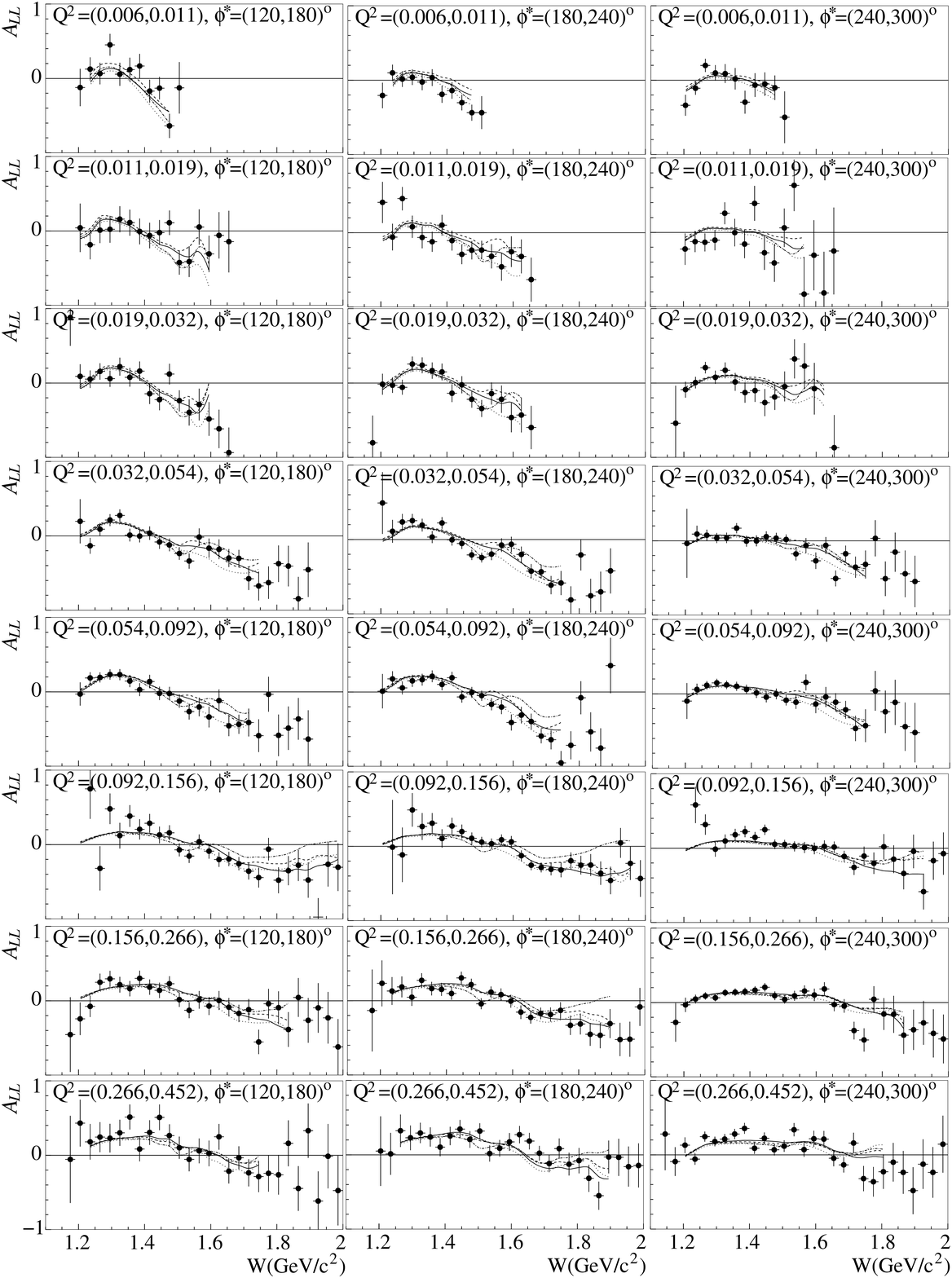}
 \caption{Results on the double-spin symmetries $A_{LL}$ for the $\vec e\vec p\to e\pi^+(n)$ channel 
   as a function of the invariant mass $W$ in GeV$/c^2$, integrated over $\cos\theta^*=(0.5,1.0)$,
   for increasing $Q^2$ ranges and three 60$^\circ$ $\phi^*$ bins. 
From top to bottom the $Q^2$ bins are
$(0.00646,0.011)$ and $(0.011,0.0187)$ (1.1 GeV NH$_3$ long bottom target), 
$(0.0187,0.0317)$ and 
$(0.0317,0.054)$ (1.3~GeV NH$_3$ long top target), 
$(0.054,0.0919)$ (2.0~GeV NH$_3$ long top target), 
$(0.0919,0.156)$, 
$(0.156,0.266)$, and
$(0.266,0.452)$~(GeV/$c$)$^2$ (3.0~GeV NH$_3$ long top target).
From left to right the $\phi^*$ bins are
$\phi^*=(120^\circ,180^\circ)$, 
$(180^\circ,240^\circ)$ and 
$(240^\circ,300^\circ)$. 
In each panel, the horizontal scale is from 1.1 to 2~GeV$/c^2$ in $W$ and the vertical scale is 
from $-1$ to $1$. 
Data are compared to four calculations: 
MAID2007 (solid)~\cite{Drechsel:2007if}, JANR (dashed)~\cite{Aznauryan:2009mx},
SAID (dash-dotted)~\cite{Arndt:2009nv}, 
and DMT2001 (dotted)~\cite{Kamalov:1999hs}.
}\label{fig:aet_w}
\end{figure}
\end{widetext}

\end{document}